\documentclass[a4paper,11pt]{article}
\usepackage[utf8]{inputenc}
\usepackage[english]{babel}
\usepackage{comment}
\usepackage{braket}
\usepackage{dsfont}
\usepackage{bbm}
\usepackage{mathdots}
\usepackage{stackrel}
\usepackage{xcolor}
\usepackage{mathtools}
\usepackage{bbm}
\usepackage{subfigure}
\usepackage[T1]{fontenc}               
\usepackage[left=3cm,
            right=2.5cm,
            top=2.5cm,
            bottom=3cm
            ]{geometry}                
\usepackage{graphicx}
\usepackage{appendix}
\usepackage{fancyhdr}
\usepackage{amsmath,                   
            amssymb,                   
            amsthm}                    
\usepackage[margin=30pt, bf, font=small, center, justification=justified]{caption}[2004/07/16]

\newif\ifnatbibsort\natbibsorttrue

\relax

\ifnatbibsort\RequirePackage[numbers,sort&compress]{natbib}\else\RequirePackage[numbers,compress]{natbib}\fi
\RequirePackage[colorlinks=true
,urlcolor=blue
,anchorcolor=blue
,citecolor=blue
,filecolor=blue
,linkcolor=blue
,menucolor=blue
,pagecolor=blue
,linktocpage=true
,pdfproducer=medialab
,pdfa=true
]{hyperref}

\begin{document} 

\title{\bf Symmetry-resolved entanglement in fermionic systems with dissipation}

\author{Sara Murciano$^{1,2}$, Pasquale Calabrese$^{3,4}$ and Vincenzo Alba$^5$}

\maketitle
{\small
\vspace{-5mm}  \ \\ \medskip
$^{1}$	Walter Burke Institute for Theoretical Physics, Caltech, Pasadena, CA 91125, USA\\
\medskip
$^{2}$	Department of Physics and IQIM, Caltech, Pasadena, CA 91125, USA\\
\medskip
{$^{3}$}  SISSA and INFN Sezione di Trieste, via Bonomea 265, 34136 Trieste, Italy\\
\medskip
{$^{4}$}  International Centre for Theoretical Physics (ICTP), Strada Costiera 11, 34151 Trieste, Italy\\
\medskip
  $^{5}$ 
 Dipartimento di Fisica dell'Universit\`a di Pisa and INFN, Sezione di Pisa, I-56127 Pisa, Italy
\medskip
}

\begin{abstract}
We investigate symmetry-resolved entanglement in out-of-equilibrium 
fermionic systems subject to gain and loss dissipation, which preserves the block-diagonal structure of the reduced density matrix. We derive a hydrodynamic 
description of the dynamics of several entanglement-related quantities, such as 
the symmetry-resolved von Neumann entropy and the charge-imbalance-resolved 
fermionic negativity. 
We show that all these quantities admit a hydrodynamic description in terms of 
entangled quasiparticles. 
While the entropy is dominated by dissipative processes, 
the resolved negativity is sensitive to the presence of entangled quasiparticles, 
and it shows the typical ``rise and fall'' dynamics. 
Our results hold in the weak-dissipative hydrodynamic limit of large intervals, long times 
and weak dissipation rates. 
\end{abstract}

\section{Introduction}
\label{sec:intro-0}

The study of entanglement dynamics is of fundamental importance to understand 
the process of thermalisation or the lack thereof in out-of-equilibrium 
quantum many-body systems~\cite{polkovnikov2011colloquium,gogolin2016equilibration,
dalessio2016quantum,calabrese2016introduction,essler2016quench}. In one-dimensional 
integrable systems the entanglement dynamics after a quantum quench
is amenable of a hydrodynamic description within the framework of the famous quasiparticle 
picture~\cite{calabrese2005evolution,fagotti2008evolution,alba2017entanglement,alba2018entanglement,bertini2022growth,calabrese2020entanglement,ac-17a}. 
Specifically, the entanglement dynamics  is 
determined  by the ballistic propagation of entangled pairs of quasiparticles. 
Importantly, recent progress with cold-atom and trapped-ion systems allows us to experimentally 
measure entanglement-related quantities~\cite{kaufman2016quantum,lukin2019probing,brydges2019probing,elben2020mixed} and their dynamics. 

The starting point for all entanglement-related quantities is the reduced density matrix $\rho_A$ of a subsystem $A$.
For example, the entanglement R\'enyi entropies are defined as  
\begin{equation}
	S_n=\frac{1}{1-n}\log(\mathrm{Tr}\rho^n_A). 
\end{equation}
The limit $n\to1$ gives the von Neumann entropy as 
\begin{equation}
	S_1=-\mathrm{Tr}\rho_A\log \rho_A. 
\end{equation}
The R\'enyi entropies and the von Neumann entropy are proper entanglement 
measures for bipartite pure states only. If the system is in a bipartite 
mixed state, neither the von Neumann entropy nor the mutual information are 
\emph{bona fide} entanglement measures. In these situations, for fermionic 
systems, one can use the 
logarithmic negativity or the fermionic logarithmic negativity 
(see section~\ref{sec:def} for precise definitions). 

Recently, there has been growing interest in the interplay between 
symmetries and entanglement. Specifically, if the system possesses a global 
symmetry, for instance, particle number conservation, the reduced density matrix 
exhibits a block structure, each block corresponding to a different symmetry sector~\cite{Laflorencie_2014,Goldstein_2018,sierra}. 
The question how the different symmetry sectors contribute to the entanglement entropies 
attracted a lot of attention, also experimentally~\cite{lukin2019probing,rvm-22,vitale2022symmetry,neven2021symmetry}. 
On the theoretical side, several works focused on the symmetry-resolved 
entropies, especially in field theory \cite{Goldstein_2018,sierra,mdgc-20, hc-20, hcc-21, hcc-a-21, chcc-a22, ccadfms-22, ccadfms-22-2, cmca-23, mca-23,Capizzi-Cal-21,Hung-Wong-21,cdm-21,mt-22,Ghasemi-22,amc-22-2,fmc-22,dgmnsz-22,cmc-23,znm-20,wznm-21,znwm-22,bbcg-22,eim-d-21} and spin chains both at equilibrium~\cite{Bonsignori_2019,Fraenkel_2020, ms-21,tr-19,ccgm-20,Murcianobis_2020,Barghathi_2018,Barghathi_2019,Barghathi_2020,wv-03,mrc-20,trac-20,kusf-20,kusf-20b,kufs-21,kufs-21-1,zshgs-20,amc-22,boncal-21,jones-22,mcp-22}, as well as in out-of-equilibrium situations~\cite{gilles_PRB,parez2021exact,fraenkel2021entanglement,Scopa_2022,pvcc-22,amc-22-3,bcckr-22,amvc-23}. 
The charge-imbalance-resolved negativity has been investigated at equilibrium~\cite{Cornfeld_2018,Murciano_2021,Chen-22}, 
and also its out-of-equilibrium dynamics was addressed~\cite{Feldman_2019,parez2022dynamics,chen-22-2}. 
Remarkably, the resolved negativity can be accessed 
experimentally~\cite{neven2021symmetry,vitale2022symmetry}. 

Still, most of the effort so far focussed on closed quantum many-body systems. 
Since in generic systems the interaction with an environment is unavoidable, 
it is crucial to understand the behaviour of  entanglement and 
symmetry-resolved entanglement 
in out-of-equilibrium \emph{open} quantum many-body systems. This is in general a 
challenging task, and results are available only in few situations, for instance, 
within the simplified framework of the Lindblad master equations~\cite{petruccione2002the}. 
For quadratic Lindblad master equations, i.e., for free-fermion and free-boson models with 
linear Lindblad operators, the out-of-equilibrium dynamics of the R\'enyi entropies, and 
of the mutual information can be understood within the  
quasiparticle picture~\cite{alba2021spreading,carollo2022dissipative,alba2022hydrodynamics}. 
This result holds in the weak-dissipative hydrodynamic  limit of 
long times and large subsystems, and weak dissipation. 
Moreover, some results are available 
for the fermionic logarithmic negativity as well. For instance, the dynamics of the 
negativity in free-fermion systems subject to gain and loss 
dissipation was investigated in Ref.~\cite{alba2022logarithmic}. Interestingly, 
due to the mixedness of the global state, the negativity 
is not related to the R\'enyi mutual information, in contrast with the unitary 
case~\cite{alba2019quantum}. 
The dynamics of both the von Neumann entropy and the negativity in the presence 
of localised losses has been studied recently~\cite{alba2021unbounded,caceffo2023entanglement}. 
Relatedly, it has been shown that 
in free-fermion models in contact with localised thermal baths, which are treated 
within the Lindblad formalism, the mutual information 
exhibits a logarithmic scaling~\cite{dabbruzzo2022logarithmic}.  
Finally, symmetry-resolved entanglement in 
open quantum many-body systems has received comparatively little attention, 
although some numerical and perturbative results are available~\cite{vitale2022symmetry}. 

Here we build a quasiparticle picture 
for several symmetry-resolved entanglement measures in free fermions subject to gain and 
loss dissipation. Such dissipation shall violate the particle number conservation, 
i.e. the $U(1)$ symmetry. However, as we will discuss below, 
the local Lindblad jump operators, which define the dissipative mechanism, 
preserve the block-diagonal structure of the reduced density matrix. We 
can refer to this scenario as a weak $U(1)$ 
symmetry~\cite{albert,prosen,vitale2022symmetry,Macieszczak,mamn-21,prosen2}.
 Specifically, we provide analytic results for both 
the symmetry-resolved von Neumann entropy,  
and the charge-imbalance-resolved fermionic negativity. 
Similar  to the non-resolved von Neumann entropy, the 
symmetry-resolved one is dominated by volume-law terms, which originate 
from dissipation. These are accompanied by terms that retain information about 
entangled quasiparticles. Precisely, these terms have the same structure as in 
the non-dissipative case. However, the entanglement content of the correlated 
quasiparticles is renormalised by the dissipative processes, and it vanishes 
exponentially at long times, reflecting how dissipation destroys genuine quantum correlation. 
While the entropies are dominated by dissipative contributions, we show that the 
fermionic symmetry-resolved negativity is determined by the entangled quasiparticles. 
Indeed, the resolved negativity exhibits the typical ``rise and fall'' 
dynamics, i.e., it grows at short times $t\lesssim 1/\gamma^\pm$, with $\gamma^\pm$ 
the dissipation rates, and it vanishes at long times. Interestingly, for generic 
gain/loss processes the 
symmetry-resolved entropies do not exhibit a time delay,  meaning that 
they are different from zero for any $t>0$, as already numerically observed  
in~\cite{vitale2022symmetry}. This reflects that different 
charge sectors are immediately ``populated'' due to the presence of the dissipation. 
This represents a crucial difference with respect to the case without dissipation~\cite{parez2021exact}.

%
\begin{figure}
\begin{center}
\includegraphics[width=0.75\linewidth]{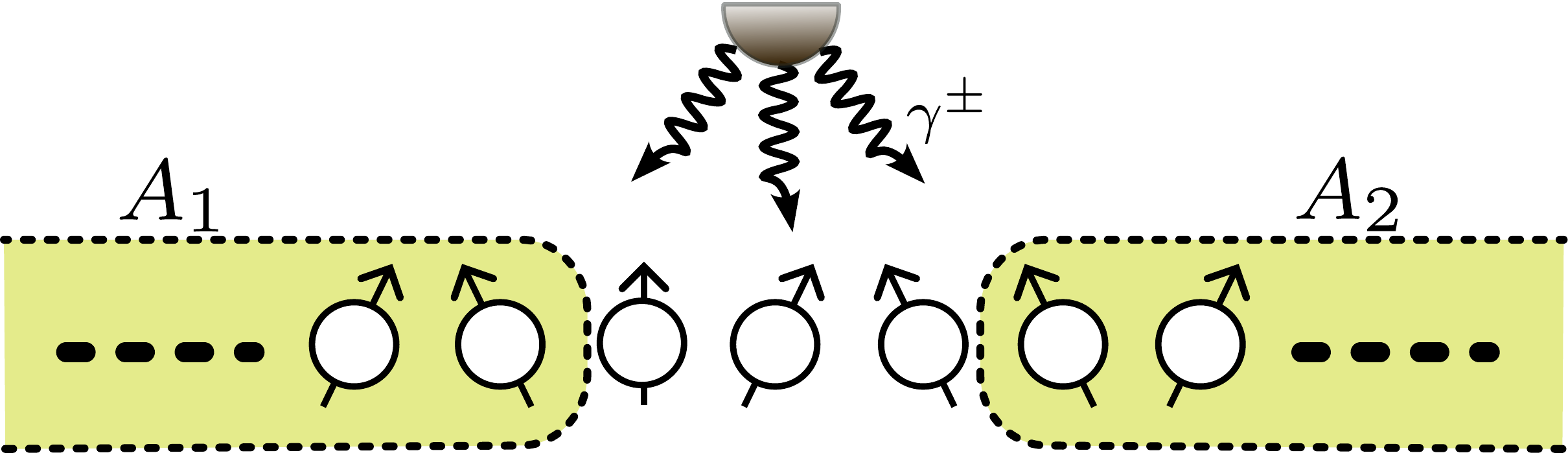}
\caption{ Schematic view of our setup. A fermionic chain  is subject to 
 gain and loss dissipation: Fermions are added or removed incoherently 
 at rates $\gamma^\pm$. The incoherent processes act on each site of the 
 chain independently. The chain is initially prepared in a low-entangled 
 state $|\Psi_0\rangle$. Here we choose as initial states the fermionic 
 N\'eel state $|\mathrm{N}\rangle$ and the Majumdar-Ghosh state 
 $|\mathrm{D}\rangle$. The chain evolves under the action of the 
 tight-binding fermionic Hamiltonian and the Lindblad operators. 
 We are interested in the symmetry-resolved entanglement entropy of 
 a subregion $A$ with the rest. We also consider the charge-imbalance-resolved 
 negativity between two equal intervals $A_1$ and $A_2$ of 
 length $\ell$ and placed at distance $d$. 
}
\label{fig:cartoon}
\end{center}
\end{figure}
%

The manuscript is organised as follows. We start in section~\ref{sec:intro} reviewing 
quantum quenches in free-fermion models without dissipation. 
In section~\ref{sec:lin} we introduce the 
dissipation and its treatment within the framework of the Lindblad master equation. 
In section~\ref{sec:def} we introduce the symmetry-resolved von Neumann entropy and 
the resolved negativity, discussing also their calculation 
in free-fermion models. Section~\ref{sec:results} 
is devoted to the presentation of our main results. Specifically, in section~\ref{sec:res-1} we 
discuss the charged moments of the reduced density matrix, which are essential 
ingredients to compute the symmetry-resolved entropy. In section~\ref{sec:quadratic} we discuss 
how to obtain the symmetry-resolved entropy and several useful 
approximations. In section~\ref{sec:quasi-neg} we present our results for 
the charge-imbalance-resolved negativity. In particular, 
we discuss the charged moments of the partial time-reversed 
reduced density matrix, and the resolved negativity. 
In section~\ref{sec:num} we provide numerical benchmarks. Specifically, 
in section~\ref{sec:ben-mom} and 
section~\ref{sec:ben-ent} we discuss the charged moments and the symmetry-resolved entropies, 
respectively. In section~\ref{sec:qua} we address the validity of the 
quadratic approximation for the charged moments. Finally, in section~\ref{sec:log-corr} we 
discuss the scaling in the weak-dissipative hydrodynamic limit. 
In particular, we highlight the presence of logarithmic corrections, 
which arise from the charge fluctuations between the subsystem and the 
rest. These are revealed in the scaling of the number entropy. 
In section~\ref{sec:ben-neg} we benchmark our results for the resolved 
negativity. We present our conclusions in section~\ref{sec:concl}.

\section{Quantum quenches in free-fermion models}
\label{sec:intro}

We consider the fermionic chain defined by the Hamiltonian 
\begin{equation}
	\label{eq:ham}
	H=\frac{1}{2}\sum_{j=1}^{L}(c^\dagger_j c_{j+1}+c^\dagger_{j+1}c_j). 
\end{equation}
Here $L$ is the chain size, $c_j^\dagger$ and $c_j$ are canonical fermionic 
creation and annihilation 
operators with anticommutation relations $\{c^\dagger_j,c_l\}=\delta_{jl}$ and $\{c_j,c_l\}=0$. 
The Hamiltonian~\eqref{eq:ham}  is diagonalised by going to Fourier space defining the fermionic 
operators $b_k:=\frac{1}{\sqrt{L}}\sum_j e^{ik j} c_j$, with the quasimomentum $k=2\pi p/L$ and $p=0,1,\dots,L-1$. The Hamiltonian~\eqref{eq:ham} becomes diagonal as 
\begin{equation}
	\label{eq:ham-diag}
	H=\sum_k \varepsilon(k) b_k^\dagger b_k,\quad\mathrm{with}\,\,\varepsilon(k):=\cos(k). 
\end{equation}
Here we defined the single-particle energy levels $\varepsilon(k)$. 
In the following we are considering the thermodynamic limit $L\to\infty$. 
For later convenience  
we define the group velocity of the fermionic excitations $v(k)$ as 
\begin{equation}
\label{eq:g-v}
v(k):=\varepsilon'(k)=d\varepsilon(k)/dk. 
\end{equation}
We focus on the nonequilibrium dynamics after the 
quench from the fermionic N\'eel state $|\mathrm{N}\rangle$ and 
Majumdar-Ghosh (dimer) state $|\mathrm{D}\rangle$, defined as 
\begin{align}
	\label{eq:neel}
	&|\mathrm{N}\rangle:=	\prod_{j=1}^{L/2}c^\dagger_{2j}|0\rangle\\
	\label{eq:dimer}
	&|\mathrm{D}\rangle:= 	\prod_{j=1}^{{L/2}}\frac{c^\dagger_{2j}-c^\dagger_{2j-1}}{\sqrt{2}}|0\rangle, 
\end{align}
with $|0\rangle$ the fermionic vacuum state. 

Let us  now discuss the quench protocol. 
At time $t=0$ we prepare the system in $|\mathrm{N}\rangle$ or $|\mathrm{D}\rangle$. 
At $t>0$ the chain undergoes unitary dynamics under the Hamiltonian~\eqref{eq:ham}. 
The fermionic correlation function $\widetilde C_{jl}$ is the central object to address 
entanglement related quantities in free-fermion systems~\cite{peschel2009reduced}. 
The matrix $\widetilde C_{jl}$ is defined as 
\begin{equation}
	\label{eq:c-corr}
	\widetilde C_{jl}:=\langle c^\dagger_j c_l\rangle. 
\end{equation}
Here $\langle\cdot\rangle$ denotes the expectation value calculated with  
the time-dependent wavefunction. 
The dynamics of the correlation function 
$\widetilde{C}_{jl}(t)$ (cf.~\eqref{eq:c-corr}) after the quench from both the N\'eel state 
and the Majumdar-Ghosh state can be derived analytically (see for instance, Ref.~\cite{parez2022dynamics}). For the N\'eel state it is straightforward to obtain  
\begin{equation}
	\label{eq:neel-c}
	\widetilde{C}_{jl}=\frac{1}{2}\delta_{jl}+\frac{1}{2}(-1)^l\int_{-\pi}^\pi\frac{dk}{2\pi}e^{ik(j-l)+2it\varepsilon(k)}=\frac{1}{2}\delta_{jl}+\frac{1}{2}(-1)^l J_{j-l}(2t),  
\end{equation}
where $J_x(t)$ is the Bessel function of the first kind. In~\eqref{eq:neel-c} is to 
stress that the dynamics is unitary. 

It is useful to exploit the invariance of the N\'eel state and the Majumdar-Ghosh state 
under translation by two sites. Thus, we rewrite~\eqref{eq:c-corr} as 
\begin{equation}
	\label{eq:neel-c-1}
\left(
\begin{array}{cc}
	\widetilde C_{2j,2l} &\widetilde C_{2j,2l-1}\\
	\widetilde C_{2j-1,2l} & \widetilde C_{2j-1,2l-1}
\end{array}
\right):=\left\langle\left(\begin{array}{c}
		c_{2j}^\dagger\\
		c_{2j-1}^\dagger 
\end{array}\right)\left(\begin{array}{cc}
		c_{2l}& c_{2l-1}
\end{array}\right)\right\rangle=
\int_{-\pi}^\pi\frac{dk}{2\pi}e^{2ik(j-l)} \hat t(k), 
\end{equation}
where now $j,l$ label the position of a two-site unit cell and $\hat t(k)$ is a 
two-by-two matrix. 
The factor $2$ in the exponent in the integral in~\eqref{eq:neel-c-1} reflects translation 
invariance by two sites. From~\eqref{eq:neel-c}, 
one obtains the matrix $\hat t(k)$ for the N\'eel state as 
\begin{equation}
	\label{eq:tk}
	\hat t_\mathrm{N}(k)=\frac{1}{2}\left(
\begin{array}{cc}
	1+e^{2it\varepsilon(k)} & e^{2it\varepsilon(k)+ik}\\
	-e^{2it\varepsilon(k)-ik} & 1-e^{2it\varepsilon(k)}
\end{array}
\right). 
\end{equation}
For the Majumdar-Ghosh state we use 
 $\hat t_\mathrm{D}(k)$ given by~\cite{f-14,parez2021exact}   
\begin{equation}
\label{eq:tk-dimer}
\hat t_\mathrm{D}(k)=\frac{1}{2}\begin{pmatrix}
 &f_+(k,t) & -g^*(k,t)\\
 &-g(k,t) & f_-(k,t)
 \end{pmatrix}
 \end{equation}
with
\begin{align}
	 & f_\pm(k,t)=1\pm\sin(k) \sin (2\varepsilon(k)t)\\
	 & g(k,t)=e^{-i k}(\cos(k) +i\sin(k)\cos(2\varepsilon(k)t)), 
\end{align}
where $\varepsilon(k)$ is given in~\eqref{eq:ham-diag}.

\section{Lindblad evolution in the presence of gain and loss  dissipation}
\label{sec:lin}

Let us now discuss the out-of-equilibrium dynamics in the tight-binding 
chain (cf.~\eqref{eq:ham})  with fermionic gain and loss processes. To describe 
these dissipative processes, we employ the 
formalism of quantum master equations~\cite{petruccione2002the}. 
Precisely, the Lindblad equation describes the time-dependent  
density matrix $\rho_t$ of the full system  as  
\begin{equation}
\label{eq:lind}
\frac{d\rho_t}{dt}=\mathcal{L}(\rho_t):=-i[H,\rho_t]+\sum_{j=1}^L\sum_{\alpha=\pm}\left(L_{j,\alpha}\rho_t 
L_{j,\alpha}^\dagger-\frac{1}{2}\left\{L_{j,\alpha}^\dagger L_{j,\alpha},\rho_t\right\}\right)\, .
\end{equation}
Here, $\{x,y\}$ is the anticommutator. The dissipation is modelled by  
$L_{j,\alpha}$, which are  known as Lindblad jump operators.  For 
gain and loss processes they are defined as  
$L_{j,-}:=\sqrt{\gamma^-}c_j$ and $L_{j,+}:=\sqrt{\gamma^+}c_j^\dagger$, 
with $\gamma^\pm$ the gain and loss rates.  
As it is clear from Eq.~\eqref{eq:lind},  incoherent absorption and emission 
of fermions happen independently at each site of the chain. This choice of the jump operators, $L_{j,\pm}$, falls into the category of weak symmetries, i.e. the generator of the master equation as a whole commutes with the symmetry operation, while each of the jump operator does not commute individually with the symmetry~\cite{prosen}. Their main feature is that they preserve the block-diagonal form of the reduced density matrix. Indeed, if the system is prepared in an initial symmetric state (like $\ket{N}$ or $\ket{D}$), 
the system state remains supported in the same symmetric eigenspace (with a given charge) when no jumps take place. On the other hand, when the first jump $L_{j_1,\pm}$ occurs at time $t_1$, the system is transformed to another eigenspace with different charge. If we interpret Eq.~\eqref{eq:lind} as a set of quantum trajectories, each one corresponding to the solution of a stochastic Schr\"odinger equation, the quantum trajectories are symmetric at all times: a single quantum jump only changes the number of particles by an integer number, but the total number of particles is conserved along each trajectory~\cite{Macieszczak}. 

For quadratic fermionic Hamiltonians, it is straightforward to obtain from~\eqref{eq:lind} 
the dynamics of  the fermionic two-point function $C_{jl}=\langle c^\dagger_j c_l\rangle=
\mathrm{Tr}(\rho_t c^\dagger_j c_l)$. Indeed,  $C(t)$ is given by~\cite{carollo2022dissipative} 
\begin{equation}
C(t)=e^{t \Lambda}C(0)e^{t \Lambda^\dagger} +\int_0^tdz 
\, e^{(t-z)\, \Lambda}\Gamma^+e^{(t-z)\, \Lambda^\dagger}\, .
\label{eq:CovMat}
\end{equation}
Here $\Lambda=i h-1/2(\Gamma^++\Gamma^-)$, where  for the tight-binding chain 
we have  $h_{jl}=1/2(\delta_{j+1,l}+\delta_{j,l+1})$.  
In~\eqref{eq:CovMat},  $\Gamma^\pm$ are $L\times L$ matrices 
defined as $\Gamma^\pm_{jl}=\gamma^\pm\delta_{jl}$. 
Since $\Gamma_{jl}^\pm$ are diagonal, $C_{jl}$ can be rewritten in 
terms of the correlation matrix $\widetilde C_{jl}(t)$ describing the dynamics 
in the absence of dissipation, i.e., with $\gamma^\pm=0$ (cf.~\eqref{eq:c-corr}). 
Precisely, one has    
\begin{equation}
	\label{eq:C-map}
	C_{jl}=n_\infty(1-b)\mathds{1}_{jl}+b \widetilde C_{jl}, 
	\quad b:=e^{-(\gamma^++\gamma^-)t},\,n_\infty:=\frac{\gamma^+}{\gamma^++\gamma^-}, 
\end{equation}
where $\mathds{1}_{jl}$ is the identity matrix element. 
Moreover, Eq.~\eqref{eq:C-map} suggests that it is convenient 
to modify the matrix $\hat t(k)$ (cf.~\eqref{eq:neel-c-1}) 
defining $\hat t'(k)$ as 
\begin{equation}
	\label{eq:tp}
	\hat t'(k)=n_{\infty}(1-b)\mathds{1}_2+b  \hat t(k), 
\end{equation}
where $\mathds{1}_2$ is the $2\times 2$ identity matrix, and $\hat t(k)$ is 
the original matrix defined in~\eqref{eq:neel-c-1}. 
The correlator $C_{jl}$ is then obtained as 
\begin{equation}
	\label{eq:diss-symb}
	C_{jl}=\int_{-\pi}^\pi\frac{dk}{2\pi} e^{2ik(j-l)}\hat t'(k). 
\end{equation}
Here $\hat t'(k)$ is given in~\eqref{eq:tp}, with $\hat t(k)$ defined 
in~\eqref{eq:tk} and~\eqref{eq:tk-dimer}.

\section{Symmetry-resolved entanglement}
\label{sec:def}

In this section we overview the definitions of 
our quantities of interest, namely 
the symmetry-resolved entanglement entropies and the charge-imbalance-resolved 
negativity. 
We focus on quantum systems with an internal  $U(1)$ symmetry 
generated by a local operator $Q$. For the tight-binding chain 
in~\eqref{eq:ham} this corresponds to the conservation of the 
number of fermions. The $U(1)$ symmetry implies that $[\rho,Q]=0$.
If the system is  bipartite in two complementary subsystems $A$ and $B$, 
since the operator $Q$ is sum of local terms,  $Q$ is decomposed as 
the sum of operators acting separately on the 
degrees of freedom of $A$ and $B$, i. e.,  
$Q=Q_A+Q_B$. By taking the trace over $B$ in $[\rho,Q]=0$ we obtain that 
$[\rho_A,Q_A]=0$, implying  that the reduced density matrix $\rho_A$ 
is block diagonal.  The different blocks correspond to a different charge 
sector with eigenvalue $q$ of $Q_A$. We have
\begin{equation}
    \label{reducedblock}
    \rho_A=\oplus_q \Pi_q\rho_A= \oplus_q[p(q)\rho_A(q)], 
\end{equation}
where $\Pi_q$ projects over the eigenspace with 
eigenvalue $q$ and we defined 
$p(q)=\mathrm{Tr} \left(\Pi_q \rho_A \right)$ as 
the probability that $Q_A$ has value $q$.
Here we normalise $\rho_A(q)$ such that $\mathrm{Tr}{\rho_A}(q)=1$. 
In the following section we introduce several entanglement-related quantities that 
can be defined from $\rho_A(q)$. This allows to understand how the 
different charge sectors of the reduced density matrix 
contribute to the entanglement. We stress again that this block-diagonal structure of $\rho_A$ is preserved by the dissipative evolution in Eq.~\eqref{eq:lind} within our choice of $L_{j,\pm}$ \cite{prosen,albert,vitale2022symmetry}.

\subsection{Symmetry-resolved entropies}
\label{sec:sr-entropies}

Let us introduce the symmetry-resolved R\'enyi and von Neumann entropy. 
The latter is the entropy calculated from  the different charge sectors of the reduced density matrix as
\begin{equation}
    \label{SymResEnt}
    S_{1}(q):=  -\mathrm{Tr} [\rho_A(q)\log(\rho_A(q))], 
\end{equation}
where, again, $\rho_A(q)$ is the normalised block of $\rho_A$ corresponding 
to the charge $q$. 
To proceed, we  plug Eq.~\eqref{reducedblock} into 
the definition of the von Neumann entropy, to obtain  the decomposition as 
\begin{equation}
\label{eq:decompositionSvN}
S_{1}=\sum_q p(q)S_{1}(q)-\sum_q p(q)\log (p(q)):=  S^{(c)}+S^{(\mathrm{num})}, 
\end{equation}
where $p(q)$ is the probability of 
measuring charge $q$ in the subsystem. 
The first term in~\eqref{eq:decompositionSvN} quantifies the average of the von Neumann entropy 
of each charge sector, and it is called 
\textit{configurational entropy}\cite{lukin2019probing,wv-03,Barghathi_2018}. 
On the other hand, $S^{(\mathrm{num})}$ measures the charge fluctuations between subsystem $A$ and 
its complement, and it is called 
\textit{number entropy}~\cite{lukin2019probing,kusf-20,kusf-20b,kufs-21-1,kufs-21,zshgs-20}. 

Notice that from $\rho_A(q)$ one can define  
the \textit{symmetry-resolved R\'enyi entropies}~\cite{Bonsignori_2019} as
\begin{equation}
    \label{SymResReny}
    S_n(q):= \frac{1}{1-n}\log \mathrm{Tr} [\rho^n_A(q)].
\end{equation} 
Here, however, we only consider the von Neumann entropy. 

We now discuss how to compute the symmetry-resolved entropy. 
In general, its computation, for instance 
in a field-theory setup, requires the spectrum 
of the symmetry-resolved reduced density matrix. This is challenging to 
extract because the projector $\Pi_q$ is a nonlocal operator. 
Here we follow a different strategy, which was put forward in 
Ref. \cite{Goldstein_2018,sierra}, and which we now outline. 
The central objects of this 
approach are the {\it charged moments}, $Z_r(\alpha)$. They are defined as 
\begin{equation}
	Z_r(\alpha)=\mathrm{Tr}[\rho_A^r e^{i\alpha Q_A}]. 
\end{equation}
By performing the Fourier transform of the charged moments, we get 
the moments of the reduced density matrix in a given charge sector, i.e., 
\begin{equation}
	\label{eq:FT}
\mathcal{Z}_r(q):=  \mathrm{Tr}(\Pi_q\rho^r)=\int_{-\pi}^{\pi}\frac{d\alpha}{2\pi}e^{-iq\alpha}Z_r(\alpha),\quad p(q) =  \mathcal{Z}_1(q),
\end{equation}
where $\Pi_q$, again, is the projector, which selects the block of the reduced density 
matrix that corresponds to value of the charge $q$. 
The symmetry-resolved R\'enyi entropies $S_r(q)$ with R\'enyi index $r$ are 
defined as 
\begin{equation}
\label{eq:sqexact}
S_r(q)=\frac{1}{1-r}\log\left(
\frac{\mathcal{Z}_r(q)}{\mathcal{Z}_1^r(q)}\right),\quad S_1(q)= \lim_{r \to 1}S_r(q),
\end{equation}
where the symmetry-resolved von Neumann entropy $S_1$ is obtained as usual 
by taking the limit $r\to1$ of $S_r$. 
Finally, for the free-fermion Hamiltonian~\eqref{eq:ham} one can write the 
charged moments in terms of the two-point fermionic correlation function as~\cite{Goldstein_2018}
\begin{equation}
	\label{eq:charged}
	\log(Z_r(\alpha))=\mathrm{Tr}\log[C_A^re^{i\alpha}+(1-C_A)^r], 
\end{equation}
where $C_A$ (cf. Eq.~\eqref{eq:c-corr}) is the fermionic correlation matrix restricted to 
part $A$ of the chain. 

\subsection{Charge-imbalance-resolved fermionic negativity}
\label{sec:def-neg}

Let us now consider the logarithmic negativity. We start by reviewing  the 
definition of the partial transpose and its relation to the time-reversal transformation. 
Let us consider a density matrix $\rho_A$ in which $A$ is further 
partitioned  as $A = A_1 \cup A_2$ (see Fig.~\ref{fig:cartoon}), where $A_1$ ($A_2$) is an interval of length $\ell_1$ ($\ell_2$).  
We can write 
\begin{equation}
\rho_A=\sum_{ijkl}\braket{e^1_i,e^2_j|\rho_A|e^1_k, e^2_l}\ket{e_i^1,e_j^2}\bra{e^1_k,e^2_l},
\end{equation}
where $\ket{e_j^1}$ and $\ket{e_k^2}$ are orthonormal bases for 
the Hilbert spaces of $A_1$ and $A_2$, respectively. 
The partial transpose $\rho^{T_1}_A$ of $\rho_A$ with respect to $A_1$ 
is defined as 
\begin{equation}
\label{eq:bosonic}
(\ket{e_i^1,e_j^2}\bra{e^1_k,e^2_l})^{T_1} :=  \ket{e_k^1,e_j^2}\bra{e^1_i,e^2_l}.
\end{equation}
The negativity is written in terms of the negative eigenvalues of $\rho_A^{T_1}$, and 
it is essentially the trace norm of $\rho_A^{T_1}$. 
In terms of its eigenvalues $\lambda_i$, the trace norm of $\rho_A^{T_1}$ can be rewritten as
\begin{equation}
\mathrm{Tr}|\rho_A^{T_1}|=\sum_i |\lambda_i|=
\sum_{\lambda_i >0}|\lambda_i|+\sum_{\lambda_i <0}|\lambda_i|=1+2\sum_{\lambda_i<0}|\lambda_i|,
\end{equation}
where in the last equality we used the normalisation $\sum_i \lambda_i=1$. 
Moreover, in the absence of negative eigenvalues, 
$\mathrm{Tr}|\rho_A^{T_1}|=1$ and the negativity vanishes.
For a bosonic system, it is known \cite{s-00} that the partial 
transpose is equivalent to the time-reversal partial transpose. 
This allows one to calculate the negativity in terms of the covariance matrix 
in bosonic systems, such as the harmonic chain~\cite{aepw-02}.  

For fermionic systems, the standard negativity is not easily computed in terms of the 
two-point correlation function, because the partial transpose is not 
a Gaussian operator~\cite{ez-15,ctc-15b}. On the other hand, the fermionic negativity \cite{ssr-17}, which 
is defined from the time-reversed partial transpose, can be computed. 
To introduce the fermionic time-reversed partial transpose it is convenient to introduce 
Majorana fermion operators $w_j$ as 
\begin{equation}
	w_{2j-1}=c_j+c^\dagger_j, \quad w_{2j}=i(c_j-c^\dagger_j), 
\end{equation}
with $c_j$ the original fermions. 
The density matrix in the Majorana representation takes the form
\begin{equation}\label{eq:dm1}
\rho_A=\sum_{\substack{\kappa,\tau \\ |\kappa|+|\tau|=
\mathrm{even}}} M_{\kappa,\tau} w_{m_1}^{\kappa_{m_1}} \dots 
w_{2m_{\ell_1}}^{\kappa_{2m_{\ell_1}}} 
w_{m'_1}^{\tau_{m'_1}} \dots w_{2m'_{\ell_2}}^{\tau_{2m'_{\ell_2}}}.
\end{equation}
Here, $w^0_x=\mathds{1}$, $w^1_x=w_x$, 
$\kappa_i, \tau_j \in \{0,1\}$, and $\kappa$ ($\tau$) is a 
$2m_{\ell_1}$-component vector ($2m'_{\ell_2}$) with entries  $0,1$ and 
with norm $|\kappa|=\sum_j \kappa_j$ ($|\tau|=\sum_j \tau_j$). 
The constraint on the parity of $|\kappa|+|\tau|$ is due to the 
fact that we require the density matrix to commute with the total fermion-number 
parity operator. In~\eqref{eq:dm1} $M_{\kappa,\tau}$ are complex numbers. 
Using Eq. \eqref{eq:dm1}, the partial time-reversed (TR) transpose $\rho^{R_1}_A$ 
with respect to the subsystem $A_1$ is defined by 
\begin{equation}\label{eq:dm}
\rho_A^{R_1}=\sum_{\substack{\kappa,\tau \\ 
|\kappa|+|\tau|=\mathrm{even}}} i^{|\kappa|}M_{\kappa,\tau} 
w_{m_1}^{\kappa_{m_1}} \dots w_{2m_{\ell_1}}^{\kappa_{2m_{\ell_1}}} 
w_{m'_1}^{\tau_{m'_1}} \dots w_{2m'_{\ell_2}}^{\tau_{2m'_{\ell_2}}}.
\end{equation}
The matrix $\rho_A^{R_1}$ is not necessarily Hermitian, 
and may have complex eigenvalues, although $ \mathrm{Tr}\rho_A^{R_1}=1$. 
Still the eigenvalues of the operator $[\rho_A^ {R_1} (\rho_A^{ R_1})^{\dagger}  ]$ are all 
real. Following Refs. \cite{ssr-17,ryu,paola} we define the fermionic negativity as 
\begin{equation}
	{\cal E}:=\log\mathrm{Tr}\sqrt{\rho_A^{R_1}(\rho_A^{R_1})^\dagger}. 
\label{NF}
\end{equation}
One can also define the (fermionic) R\'enyi logarithmic negativities ${\cal E}_r=\log(R_r)$ 
with $R_r$ given as 
\begin{equation}\label{eq:sup}
R_r= 
\begin{cases}
  \mathrm{Tr}(\rho_A^{R_1}(\rho_A^{R_1})^{\dagger}\dots  \rho_A^{R_1}(\rho_A^{R_1})^{\dagger}), &\quad r \quad \mathrm{even},\\
  \mathrm{Tr}(\rho_A^{R_1}(\rho_A^{R_1})^{\dagger}\dots \rho_A^{R_1}),& \quad r \quad \mathrm{odd}.
\end{cases}
\end{equation}
The negativity can be obtained as  
$\displaystyle \mathcal{E}=\lim_{r_e \to 1}  \log(R_{r_e})$,
where $r_e$ denotes an even integer~\cite{ssr-17}.

We can adapt the approach 
discussed in section~\ref{sec:sr-entropies} to the computation of the charged moments 
of the  time-reversed (TR) partial transpose obtaining 
\begin{equation}\label{eq:supalpha}
N_r(\alpha)= 
\begin{cases}
  \mathrm{Tr}(\rho_A^{R_1}(\rho_A^{R_1})^{\dagger}\dots  \rho_A^{R_1}(\rho_A^{R_1})^{\dagger}e^{i{Q}_A\alpha}), &\quad r \quad \mathrm{even},\\
 \mathrm{Tr}(\rho_A^{R_1}(\rho_A^{R_1})^{\dagger}\dots \rho_A^{R_1}e^{i{Q}_A\alpha}), & \quad r \quad \mathrm{odd}.
\end{cases}
\end{equation}
A Fourier transform allows us to derive the moments of the TR partial transpose 
${\mathcal{Z}}_{R_1,r}$ as 
\begin{equation}\label{eq:pqft}
    \mathcal{Z}_{R_1,r}(q)=\displaystyle \int_{-\pi}^{\pi}\frac{d\alpha}{2\pi }e^{-i\alpha q}N_{r}(\alpha), \quad p(q)=\displaystyle \int_{-\pi}^{\pi}\frac{d\alpha}{2\pi }e^{-i\alpha q}N_{1}(\alpha),
\end{equation}
from which we define the ratios $R_r(q)$ and the charge-imbalance-resolved negativity 
${\mathcal E}(q)$ as 
\begin{equation}\label{eq:rnq}
    R_{r}(q)=\frac{\mathcal{Z}_{R_1,r}(q)}{p^r(q)}, \qquad 
    \mathcal{E}(q)=\lim_{r_e\to 1}\log\left(\frac{{\mathcal Z}_{R_1,r_e}(q)}{p^{r_e}(q)}\right). 
\end{equation}
Let us stress that the replica limit for $R_{r_e}(q)$ 
requires an analytic continuation  $r_e \to 1$ from the even sequences 
$R_{r_e}$, whereas in the denominator we can set $r\to1$. 

Let us now discuss how to compute the charged moments $N_r(\alpha)$ in free-fermion systems. 
The covariance matrix $\Gamma$ is the central object to compute the charged moments and the 
symmetry-resolved negativity. $\Gamma$ is defined as 
\begin{equation}
	\label{eqw:gamma-neg}
	\Gamma_{jl}=\mathds{1}-2C_{jl}, \quad j,l\in A,  
\end{equation}
where $\mathds{1}$ is the identity matrix restricted to subsystem $A=A_1\cup A_2$, 
and $C_{jl}$ is the fermionic correlation matrix~\eqref{eq:C-map}. 
If the subsystem $A$ is made of two intervals as $A=A_1\cup A_2$ (see Fig.~\ref{fig:cartoon}), 
we can decompose $\Gamma$ as 
\begin{equation}
\Gamma= 
\begin{pmatrix}
\Gamma_{11}& \Gamma_{12} \\
\Gamma_{21} &  \Gamma_{22}
\end{pmatrix},
\end{equation}
where $\Gamma_{11}$ and $\Gamma_{22}$ are the reduced covariance matrices 
of the two subsystems $A_1$ and $A_2$, respectively,  
while $\Gamma_{12}$ and $\Gamma_{21}^{\dagger}$ contain the cross 
correlations between them.
By simple Gaussian states manipulations, the covariance 
matrices $\Gamma_\pm$ associated with $\rho^{R_1}_A$ and 
$(\rho^{R_1}_A)^{\dagger}$ (cf.~\eqref{eq:dm} for their definitions) are 
obtained as 
\begin{equation}\label{eq:gammapm}
\Gamma_{\pm}= 
\begin{pmatrix}
-\Gamma_{11}& \pm i \Gamma_{12} \\
\pm i\Gamma_{21} &  \Gamma_{22}
\end{pmatrix}.
\end{equation}
Here we are interested in $N_r(\alpha)$ defined as 
\begin{equation}
	\label{eq:nr-alpha}
N_{r}(\alpha)=\mathrm{Tr}[(\rho_A^{R_1} (\rho_A^{R_1})^{\dagger})^{r/2}e^{iQ_A\alpha}], 
\end{equation}
where $r$ is an even integer. We also consider $N_1(\alpha)$ defined as 
\begin{equation}
N_{1}(\alpha)= \mathrm{Tr}[\rho_A^{R_1} e^{i Q_A\alpha}].
\end{equation}
To proceed we need the covariance 
matrix $\Gamma_\mathrm{x}$ associated with the normalised composite density 
operator $\rho_{\mathrm{x}}=\rho_A^{R_1} (\rho_A^{R_1})^{\dagger}/Z_{\mathrm{x}}$. 
This is given as~\cite{parez2022dynamics}   
\begin{equation}
\label{eq:gammax}
\Gamma_{\mathrm{x}}=(\mathds{1}+\Gamma_+\Gamma_-)^{-1}(\Gamma_++\Gamma_-),
\end{equation}
where the normalisation factor is 
$Z_{\mathrm{x}}=\mathrm{Tr}(\Gamma_{\mathrm{x}})=\mathrm{Tr}(\rho_A^2)$. 

To proceed, let us define the charged R\'enyi negativities ${\cal E}_r(\alpha)$ 
(for even $r$) as 
\begin{multline}
	\label{eq:ealpha}
	{\cal E}_r(\alpha):=\log(N_r(\alpha))=-2i\alpha\ell +\mathrm{Tr}\log\left[
\left(\frac{\mathds{1}-\Gamma_\mathrm{x}}{2}\right)^{r/2}+
e^{i\alpha}\left(\frac{\mathds{1}+\Gamma_\mathrm{x}}{2}\right)^{r/2}\right]
\\+\frac{r}{2}\mathrm{Tr}\log\left[C_A^2+(\mathds{1}-C_A)^2\right], 
\end{multline}
where $N_r(\alpha)$ are defined in~\eqref{eq:nr-alpha} and we have set $\ell_1=\ell_2=\ell$. 
Here $C_A$ is the restricted fermionic correlation matrix, i.e., $C_{jl}$ with 
$j,l\in A$. 
Now, Eq.~\eqref{eq:ealpha} can be rewritten as 
\begin{multline}
\label{eq:num1}
{\cal E}_r(\alpha)=-2i\alpha \ell+\sum_{j=1}^{2\ell}\log
\left[ \left(\frac{1-\nu_j^{\mathrm{x}}}{2} \right)^{r/2}+
e^{i\alpha} \left(\frac{1+\nu_j^{\mathrm{x}}}{2} 
\right)^{r/2} \right]\\
+\frac{r}{2}\sum_{j=1}^{2\ell}\log \left[ \zeta_j^{2}+ (1-\zeta_j)^{2} \right],
\end{multline}
where $\nu_j^{\mathrm{x}}$ are eigenvalues of the matrices 
$\Gamma_{\mathrm{x}}$~\eqref{eq:gammax}, and $\zeta_j$ are 
the eigenvalues of the fermion correlation matrix $C_A$ 
introduced in the section~\ref{sec:intro}. 
The charged  negativity ${\cal E}(\alpha)$ is defined as 
\begin{equation}
	{\cal E}(\alpha):=\lim_{r\to1}{\cal E}_r(\alpha),\quad\mathrm{with}\,\, r\,\,\mathrm{even}. 
\end{equation}
Finally, let us observe that $N_1(\alpha)$ is written in terms of the eigenvalues  
$\nu^+_j$ of $\Gamma_+$~\eqref{eq:gammapm}
\begin{equation}
\label{eq:num}
\log(N_{1}(\alpha))=
-2i \alpha \ell+\sum_{j=1}^{2\ell}\log
\left[ \left(\frac{1-\nu^+_j}{2} \right)+e^{i\alpha} \left(\frac{1+\nu^+_j}{2} \right) \right].
\end{equation}
The strategy to  compute the charge-imbalance-resolved negativity is to 
perform   the Fourier transform of the quasiparticle prediction for 
$\log(N_{r}(\alpha))$ and $\log(N_{1}(\alpha))$.

\section{Quasiparticle picture for symmetry-resolved entropies }
\label{sec:results}

In this section we derive the quasiparticle picture for the symmetry-resolved 
entropies. In section~\ref{sec:res-1} we focus on the charged moments of the 
reduced density matrix, whereas in section~\ref{sec:quadratic} we present 
our results for the symmetry-resolved von Neumann entropy. In section~\ref{sec:quadratic} 
we also discuss some 
useful approximations that are needed to derive analytically the behaviour of the 
resolved entropies. 

\subsection{Charged moments of the reduced density matrix}
\label{sec:res-1}

%
\begin{figure}[t]
\centering
\includegraphics[width=0.58\textwidth]{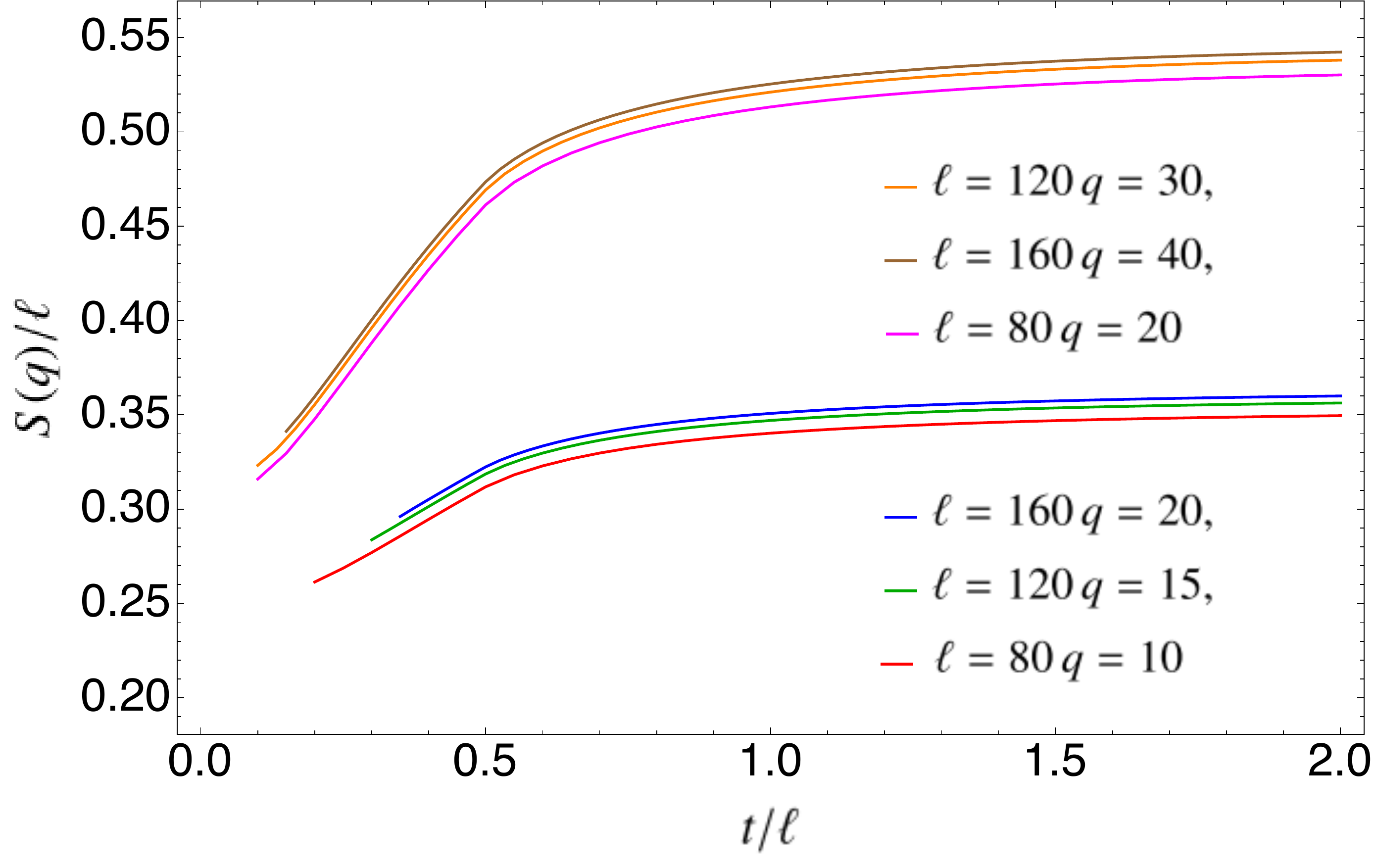}
\caption{ Effect of gain and loss dissipation in the symmetry-resolved 
 von Neumann entropy $S(q)$. We show results for the quench from the N\'eel 
 state in the tight-binding chain. We plot the density of entropy 
 $S(q)/\ell$ versus rescaled time $t/\ell$ for different charge sectors $q$. 
 Here we  fix $q/\ell=1/8$ and $q/\ell=1/4$. To ensure the 
 weak-dissipative hydrodynamic limit the dissipation rates $\gamma^\pm$ 
 are fixed as $\gamma^-=1/\ell$, $\gamma^+=5/(100\ell)$. 
}
\label{fig:sr-theo}
\end{figure}
%

Here we discuss the quasiparticle picture for the charged moments~\eqref{eq:charged} for 
the quench from the N\'eel state~\eqref{eq:neel} and dimer state \eqref{eq:dimer}. 
We consider the weakly dissipative hydrodynamic limit, which corresponds to 
$\ell,t\to\infty$, $\gamma^\pm\to0$, at fixed and arbitrary $t/\ell$ and $\gamma^\pm \ell$, with $\ell$ the length of the subsystem $A$. 

Since we are interested in the dynamics of Gaussian states, the system is 
fully characterised at any time after the quench by its two-point correlation functions. 
In the absence of pairing terms like $c^{\dagger}c^{\dagger}$ or $cc$ in the Hamiltonian 
at time $t=0$, the relevant covariance matrix is 
$C_{jl}$ (cf.~\eqref{eq:C-map}). 
Again, the correlation matrix in the presence of gain/loss dissipation is simply 
related to the dynamical correlation function $\widetilde C_{jl}$ (cf.~\eqref{eq:neel-c-1}) 
for the quench without dissipation. 
To determine the quasiparticle picture for the charged moments, it is convenient to 
employ the representation of the fermionic correlator 
in~\eqref{eq:diss-symb} and~\eqref{eq:neel-c-1}~\eqref{eq:tk}. 
To proceed, we formally expand~\eqref{eq:charged} as 
\begin{equation}
	\label{eq:Z-exp}
	\log(Z_r(\alpha))=\sum_{j=0}^\infty c_j \mathrm{Tr}\,C_A^j, 
\end{equation}
where $C_A$ is $C_{jl}$ (cf.~\eqref{eq:C-map}) with $j,l\in A$, and the coefficients $c_j$ are 
the coefficients of the Taylor expansion of $\log[x^r e^{i\alpha}+(1-x)^r]$ around 
$x=0$. 

By following the same steps of Ref.~\cite{alba2022logarithmic}, we can  find the 
quasiparticle picture for the charged moments. 
Specifically, we first consider the expansion~\eqref{eq:Z-exp}. Thus, the 
problem is reduced to that of determining the 
hydrodynamic behaviour of $\mathrm{Tr}\,C_A^j$ for arbitrary $j$. 
This last step can be performed by using the multidimensional version of the 
stationary phase approximation~\cite{wong2001asymptotic}. 
By closely following the same steps as in~\cite{alba2022logarithmic}, we obtain 
for both the quench from the N\'eel state and the dimer state that 
\begin{equation}
\label{eq:charged-quasi}
\log(Z_r(\alpha))=
\ell\int \frac{dk}{2\pi}\left[\mathrm{max}(1-2|v(k)|t/\ell,0)
h^\mathrm{mix}_r(\alpha,k) +\mathrm{min}(2|v(k)|t/\ell,1)
h^{\mathrm{q}}_r(\alpha,k)\right],
\end{equation}
where $v(k)=\sin(k)$, i.e., the same group velocity of the fermions 
in the non-dissipative case. 
In~\eqref{eq:charged-quasi} we introduced the two contributions 
$h_r^\mathrm{mix}(\alpha,k)$ and $h_r^\mathrm{q}(\alpha,k)$ defined as 
\begin{align}
	\label{eq:h-m}
	& h_r^\mathrm{mix}=\mathrm{Tr}\, h_r(\alpha,\hat t'(k)),\\
	\label{eq:h-q}
	& h_r^\mathrm{q}=h_r(\alpha,\rho_t(k)). 
\end{align}
Here the function $h_r(\alpha,x)$ is defined as 
\begin{equation}
	\label{eq:h-def}
	h_r(\alpha,x)=\log(e^{i\alpha}x^r+(1-x)^r). 
\end{equation}
The term~\eqref{eq:h-m} is of purely dissipative origin and it vanishes in 
the limit $\gamma^\pm\to0$. Precisely, $h^\mathrm{mix}_r(\alpha,k)$ corresponds to 
the charged moments obtained from~\eqref{eq:charged} by 
replacing $C_A$ with the full-system correlator. 
Notice that in~\eqref{eq:h-m} the trace is over the two-by-two matrix $\hat t'(k)$ 
(cf.~\eqref{eq:tp}). 
The term $h_r^\mathrm{q}$ in~\eqref{eq:charged-quasi} is a ``quantum'' one, 
meaning that it is determined by both the dissipative and the unitary part of the 
evolution~\eqref{eq:lind}. In the limit $\gamma^\pm\to0$ this term gives the 
result of the quasiparticle picture in the non dissipative case. 
Importantly, $h^\mathrm{q}_r(\alpha,k)$ 
depends only on the fermion density $\rho_t(k)$. In the absence of dissipation 
$\rho_t(k)$ does not depend on time, which is not the case at nonzero $\gamma^\pm$. 
For quenches in the tight-binding chain in the presence of 
gain/loss dissipation the time-evolved density 
$\rho_t(k)$ is obtained as~\cite{carollo2022dissipative} 
\begin{equation}
	\label{eq:rho-diss}
	\rho_t(k)=e^{-\gamma(k) t}\rho_0(k)+\frac{\alpha(k)}{\gamma(k)}(1-e^{-\gamma(k) t}). 	
\end{equation}
Here we have 
\begin{equation}
	\label{eq:nd-details}
	\gamma(k)=\gamma^++\gamma^-,\quad \alpha(k)=\gamma^+. 
\end{equation}
As it is clear from~\eqref{eq:nd-details} for the quench from the 
N\'eel state and the dimer state $\gamma(k)$ and $\alpha(k)$ do not 
depend on the quasimomentum $k$. This simplification holds for the 
gain/loss dissipation but it is not expected for the generic linear 
dissipation~\cite{carollo2022dissipative,alba2022hydrodynamics}. 
In~\eqref{eq:rho-diss} $\rho_0(k)$ is the density of fermionic quasiparticles 
that fully characterises the Generalised Gibbs Ensemble (GGE) in the 
absence of dissipation. Specifically, $\rho_0(k)$ is computed as 
$\rho_0(k)=\langle\psi_0|c^\dagger_k c_k|\psi_0\rangle$. 
For the quench from the N\'eel and the dimer state, a straightforward 
calculation gives the densities $\rho^\mathrm{N}_0(k)$ and $\rho^\mathrm{D}_0(k)$ as 
\begin{equation}
	\rho^\mathrm{N}_0(k)=\frac{1}{2},\quad \rho_0^\mathrm{D}(k)=
	|\cos(k)|. 
\end{equation}
Finally, by using the explicit form of $\hat t'(k)$ (cf.~\eqref{eq:tp}) for 
the N\'eel and the dimer quench, we obtain that $h^{\mathrm{mix}}_r$ 
is the same for both of them, and it reads as 
\begin{equation}
	\label{eq:mixed}
	h^\mathrm{mix}_r(\alpha,k) =\frac{h_r(\alpha,n_0)+h_r(\alpha,n_1)}{2}. 
\end{equation}
Here  $n_0$ and $n_1$ are given as 
\begin{equation}
	\label{eq:n01}
	n_0=\frac{\alpha(k)}{\gamma(k)}(1-e^{-\gamma(k)t}),\quad 
	n_1=e^{-\gamma(k)t}+\frac{\alpha(k)}{\gamma(k)}(1-e^{-\gamma(k)t}), 
\end{equation}
where $\gamma(k)$ and $\alpha(k)$ are given in~\eqref{eq:nd-details}. 
Notice that $n_0$ is obtained from~\eqref{eq:rho-diss} by setting 
$\rho_0(k)=0$ and $n_1$ by fixing $\rho_0(k)=1$. This reflects 
that at $t=0$ for both the N\'eel state and the dimer state half of the eigenvalues 
of the correlation matrix $C_{jl}$ (cf.~\eqref{eq:c-corr}) are zero and 
half are one. They do not depend on time in the absence of dissipation. 
For gain/loss dissipation, the eigenvalues evolve independently according 
to~\eqref{eq:rho-diss}. 
It is also interesting to observe that we can rewrite~\eqref{eq:charged-quasi} as 
\begin{equation}
	\label{eq:charged-quasi-1}
	\log(Z_r(\alpha))=\int\frac{dk}{2\pi} \Big\{\ell h_r^\mathrm{mix}(\alpha,k)+
	\min(2|v(k)|t,\ell)\big[h_r^\mathrm{q}(\alpha,k)-h_r^\mathrm{mix}(\alpha,k)\big]\Big\}. 
\end{equation}
From~\eqref{eq:charged-quasi-1} it is clear that $h_r^\mathrm{mix}$ plays a twofold 
role. Specifically, $h_r^\mathrm{mix}$ gives a volume-law contribution (first 
term in~\eqref{eq:charged-quasi-1}) which 
does not contain information about the dynamics of the quasiparticles. On the other hand, 
$h_r^\mathrm{mix}$ also appears with a minus sign in the second term in~\eqref{eq:charged-quasi-1}, 
which is the ``quantum'' one because it is the only one surviving in the limit $\gamma^\pm\to0$.  

Let us now discuss some limits of $\log(Z_r(\alpha))$. First, at $t=0$ 
since $h_r(\alpha,0)=0$ and $h_r(\alpha,1)=i\alpha$, one has that $\log(Z_r(\alpha))=
i\alpha\ell/2$. 
For $t\to\infty$, one has that  $n_0,n_1\to \alpha(k)/\gamma(k)=
n_\infty$ (cf.~\eqref{eq:C-map}) and $\rho_t\to n_\infty$. This implies that the 
second term in~\eqref{eq:charged-quasi-1} vanishes and finally one obtains
$\log(Z_r(\alpha))\to \ell h_r^\mathrm{mix}(\alpha,n_\infty)$.

\subsection{Computing the symmetry-resolved entropies}
\label{sec:quadratic}

To determine the quasiparticle picture for the symmetry-resolved entropies $S_r(q)$ 
(cf.~\eqref{eq:sqexact}), 
one has to employ the quasiparticle picture for the charged moments $\log(Z_r(\alpha))$ 
in~\eqref{eq:FT} and perform the Fourier transform. This is in general a challenging 
task, although closed-form expressions have been obtained in the nondissipative 
case~\cite{parez2021exact}. Clearly, the Fourier transform in~\eqref{eq:FT} can be 
performed numerically and we report the analytical predictions for the symmetry-resolved entropies in Fig.~\ref{fig:sr-theo}. However, this computation suffers from a numerical instability in solving the integral, and for this reason we omit the result for short time.

Another strategy is to use a saddle point approximation. Indeed, in~\eqref{eq:FT} one 
has to assume $q\propto \ell$ in order to have non trivial results. 
We have 
\begin{equation}\label{eq:FTintegral}
	{\mathcal Z}_r(q)=\int_{-\pi}^\pi\frac{d\alpha}{2\pi} e^{\ell h_r(\alpha)}, 
\end{equation}
where we defined $h_r(\alpha)$ as 
\begin{equation}
\label{eq:saddle}
h_r(\alpha)=-i\frac{q}{\ell}\alpha+\int_{-\pi}^\pi 
\frac{d k}{2\pi}\Big\{h^\mathrm{mix}_r(\alpha,k)
+\mathrm{min}(2|v|t/\ell,1)\big[h^\mathrm{q}_r(\alpha,k)-
h^\mathrm{mix}(\alpha,k)\big]\Big\}.
\end{equation} 
The saddle point condition for $\alpha$ reads as 
\begin{equation}
	\label{eq:saddle-1}
	\partial_\alpha h_r(\alpha)\Big|_{\alpha=\alpha^*}=0. 
\end{equation}
Since the integration over $k$ in~\eqref{eq:saddle} can be performed 
analytically, the resulting saddle point condition for $\alpha^*$ can be effectively 
solved numerically. Moreover, analytic solutions of~\eqref{eq:saddle} are 
also possible, although the obtained expressions are cumbersome, and we do 
not report them. 
Instead, we now discuss some features of the solution $\alpha^*$.

In the absence of dissipation, $\alpha^*$ is a complex number and the presence of a nonzero real part leads to a delay in the evolution of the symmetry-resolved entropies, i.e., the entropies start to be nonzero for $t>t_d$, with $t_d$ the delay time~\cite{parez2021exact}. Specifically, for each value of the charge $q$ we have that $\alpha^*$ in the 
non-dissipative situation is given as 
\begin{equation}
	\label{eq:alphas}
	\alpha^*=i\log\left(\frac{\mathcal{J}(t)-2\Delta 
	q/\ell}{\mathcal{J}(t)+2\Delta q/\ell}\right),\quad\mathrm{with}\,\, \Delta q:= q-\frac{\ell}{2}. 
\end{equation}
Here we defined 
\begin{equation}
	\label{eq:J-def}
	\mathcal{J}(t):=\int_{-\pi}^\pi\frac{dk}{2\pi}\min(2|v(k)|t,\ell). 
\end{equation}
In the presence of gain and loss dissipation the delay effect is suppressed, as it was already shown in~\cite{vitale2022symmetry}. 

For simplicity, we focus on a quench from the N\'eel state $|\mathrm{N}\rangle$. First,  the saddle point condition~\eqref{eq:saddle-1} has three 
solutions. By a standard saddle point analysis~\cite{wong2001asymptotic}, we observe that the only physical solution describing the exact lattice computations are the ones with zero real part for any time $t$. 
%
\begin{figure}[t]
\centering
{\includegraphics[width=0.6\textwidth]{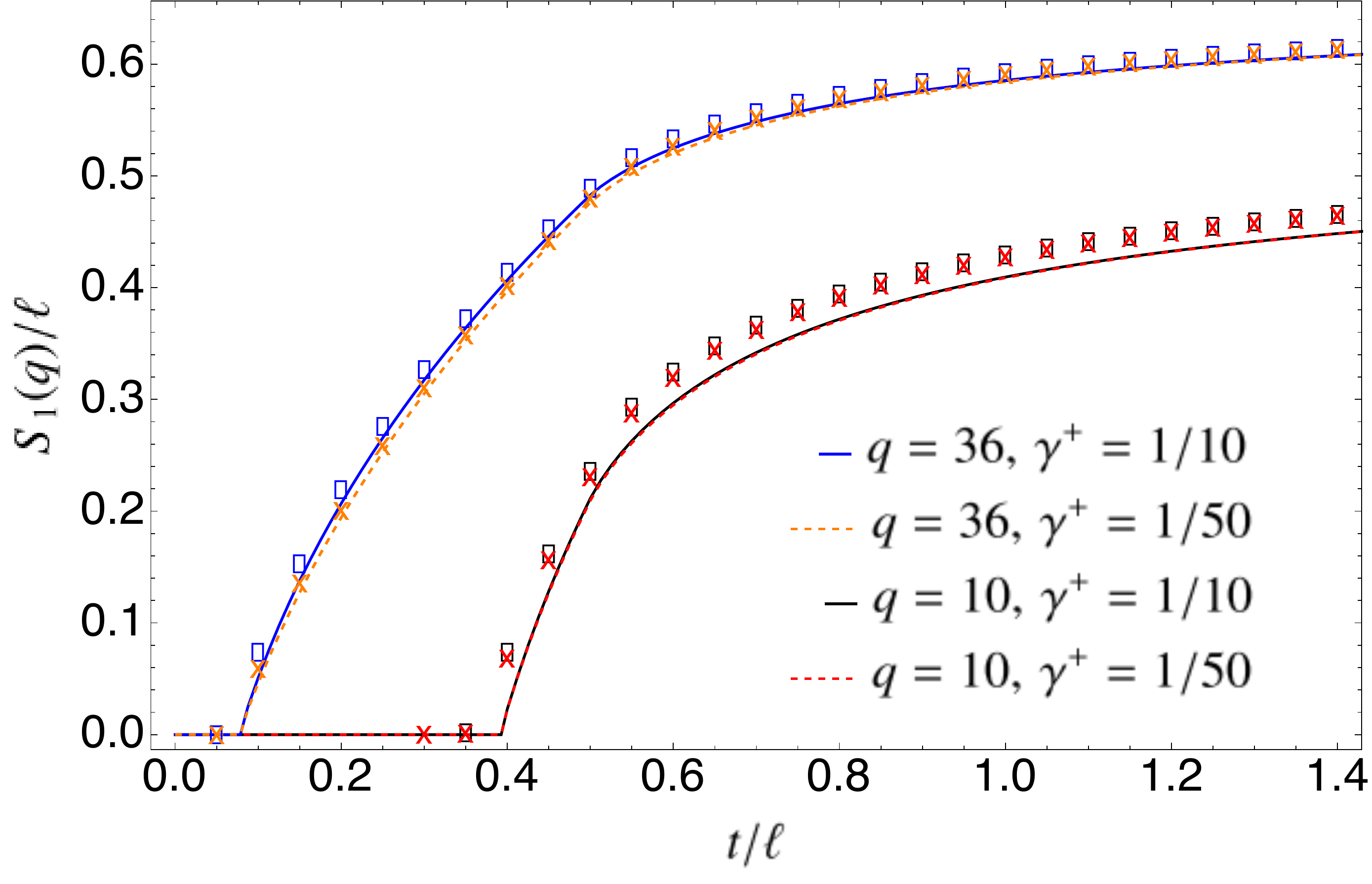}}\caption{ Time-delay in the dynamics of the symmetry-resolved von Neumann  entropy with pure gain ($\gamma^-=0$). We plot $S_1(q)/\ell$ versus the rescaled time $t/\ell$, and fixed $\ell=80$,  for the quench from the fermionic N\'eel state in the tight binding chain. The different lines are the results obtained from the quasiparticle picture for different values of the charge $q$ and dissipation $\gamma^+$, using the saddle point approximation.
For some values of $t/\ell$  we report the numerical value of the symmetry-resolved entropy obtained through exact simulations of the dynamics (symbols). We clearly observe a delay time in agreement with Eq.~\eqref{eq:saddle-n}, which does not depend on the dissipative term $\gamma^+$. }
\label{fig:tdelay}
\end{figure}
%
In the limit $\gamma^+\to 0$ ($\gamma^-\to 0$), i.e. pure loss or pure gain, we still find a delay time for $q>\ell/2$ ($q<\ell/2$). Indeed, in this case the saddle point condition has two solutions and, at short times $t \leq t_d$, the dominant solution of the saddle point condition gives a complex entropy which physically means 
that it averages to zero. As in the non-dissipative case, one finds that the delay time is obtained by solving the equation 
\begin{equation}
	\label{eq:saddle-n}
	{\mathcal J}(t)-2\frac{|\Delta q|}{\ell}=0,\quad\Delta q=q-\frac{\ell}{2}. 
\end{equation}
In Fig.~\ref{fig:tdelay} we report the behaviour of the symmetry-resolved entropy using the saddle point approximation in the presence of pure gain and we show that, for $q<\ell/2$ it displays a delay, which is also confirmed by the exact lattice computations.

Having solved, for instance numerically, the saddle-point constraint~\eqref{eq:saddle-1}, 
${\mathcal Z}_r(q)$ for $t>t_d$ is given as 
\begin{equation}
	{\mathcal Z}_r(q)=e^{\ell h_r(\alpha^*)}\sqrt{\frac{1}{2\pi\ell|h_r''(\alpha^*)|}}, 
\end{equation}
where $\alpha^*$ is the solution of~\eqref{eq:saddle-1} with $\mathrm{Re}(\alpha^*)=0$. 

Another useful strategy to gain insights into the symmetry-resolved entropies is to 
use a quadratic approximation, by expanding the charged moments around $\alpha=0$. 
One obtains 
\begin{equation}
\log(Z_r(\alpha))\simeq\log(Z_r(0))+i\alpha\mathcal{B}^{(1)}_r-\frac{\alpha^2}{2}\mathcal{B}^{(2)}_r,
\end{equation}
where 
\begin{equation}
	\mathcal{B}^{(1)}_r=\partial_{\alpha}\log(Z_r(\alpha)) \Big |_{\alpha=0},\quad
	\mathcal{B}^{(2)}_r=\partial^2_{\alpha}\log(Z_r(\alpha)) \Big |_{\alpha=0}.
\end{equation}
With this quadratic
expansion, we can perform the Fourier transform in~\eqref{eq:FT}. We obtain 
\begin{equation}
\label{eq:sp}
\mathcal{Z}_r(q)=Z_r(0)e^{-\frac{(q-\mathcal{B}^{(1)}_r)^2}{2\mathcal{B}^{(2)}_r}}\frac{1}{\sqrt{2\pi\mathcal{B}^{(2)}_r }}.
\end{equation}
This approximation is valid only until 
$\Delta q =q-\mathcal{B}^{(1)}_r $ is ${\mathcal O}(\sqrt{\mathcal{B}^{(2)}_r})$, 
or, in other words, for small fluctuations $\Delta q $. 
Within this approximation, the corresponding symmetry-resolved entropies read 
\begin{equation}
\label{eq:sqapprox}
S_r(q)=S_r-\frac{1}{2}\log(2\pi)-\frac{1}{2(1-r)}\log\frac{\mathcal{B}^{(2)}_r}{(\mathcal{B}^{(2)}_1)^r}-\frac{(q-\mathcal{B}^{(1)}_r)^2}{2(1-r)}\left[\frac{1}{\mathcal{B}^{(2)}_r}- \frac{r}{\mathcal{B}^{(2)}_1}\right]. 
\end{equation}
From Eq. \eqref{eq:sqapprox}, we can also predict the asymptotic value 
of the symmetry-resolved entropies for $t \to \infty$ at leading order in $\ell$.  
For a N\'eel state, we obtain 
\begin{multline}
	\label{eq:S1-q}
	S_1(q)=-\ell n_{\infty}\log (n_{\infty})-\ell (1-n_{\infty})\log 
		(1-n_{\infty})-1/2 \log(2\pi \ell)-1/2 \log((1-n_{\infty})n_{\infty} )
\\
+\frac{\tanh ^{-1}(1-2 n_{\infty}) \left(-2 \ell n_{\infty}^2 q+\ell n_{\infty} (n_{\infty}
(\ell+2 n_{\infty}-3)+1)+(2 n_{\infty}-1) q^2\right)}{\ell (n_{\infty}-1) n_{\infty}}\\-\frac{(\ell n_{\infty}-q)^2}{2\ell n_{\infty}(1-n_{\infty})}. 
\end{multline}
Now, Eq.~\eqref{eq:S1-q} predicts a volume-law scaling 
$\propto\ell$, and a subleading logarithmic one $\propto\log(\ell)$. 
Importantly, Eq.~\eqref{eq:S1-q} relies on the validity of the quadratic 
approximation~\eqref{eq:sqapprox}. 

It is also interesting to derive the number entropy $S^\mathrm{(\mathrm{num})}$ 
(cf.~\eqref{eq:decompositionSvN}) within the quadratic approximation. 
It is obtained from  $p(q)=\mathcal{Z}_1(q)$, and it is defined as 
\begin{equation}
	S^{(\mathrm{num})}=-\sum_q p(q) \log(p(q)). 
\end{equation}
By using the quadratic approximation for $\mathcal{Z}_1(q)$ (cf.~\eqref{eq:sp}), 
we obtain that 
\begin{equation}
\label{eq:pq}
p(q)=
\frac{1}{\sqrt{\pi}}\frac{1}{\sqrt{2\ell \rho_t
(1-\rho_t)-2\mathcal{J}'(t)(\rho_t-n_0)^2}} 
\exp\left({-\frac{(q-\ell \rho_t)^2}{2\ell \rho_t(1-\rho_t)-2\mathcal{J}'(t)
(\rho_t-n_0)^2}}\right), 
\end{equation}
where we defined $\mathcal{J}'(t)$ as 
\begin{equation}
	\mathcal{J}'(t)=\ell \int \frac{d\lambda}{2\pi}\mathrm{max}(1-2|v|t/\ell,0). 
\end{equation}
In~\eqref{eq:pq} $n_0$ and $n_1$ are the same as in~\eqref{eq:n01}, and 
$\rho_t$ is given in~\eqref{eq:rho-diss} with $\rho_0(k)=1/2$ for a N\'eel state. 
The change of variables $q \to q -\ell \rho_t$ gives 
\begin{equation}
\label{eq:snum}
S^{(\mathrm{num})}\simeq 
\int_{-\infty}^{\infty}dq p(q) \log(p(q))=
\frac{1}{2}\left[1+\log(\pi(2\ell \rho_t(1-\rho_t)-
2\mathcal{J}'(t)(\rho_t-n_0)^2) )\right].
\end{equation}
We anticipate that~\eqref{eq:snum} correctly describes the exact 
value of the number entropy.
In the large time limit, from~\eqref{eq:snum} we find
\begin{equation}
\label{eq:larget}
\lim_{t \to \infty}S^{(\mathrm{num})}
=\frac{1}{2}+\frac{1}{2}\log [2\pi \ell n_{\infty}(1-n_{\infty}) ],
\end{equation}
We will provide numerical benchmarks of~\eqref{eq:larget} in section~\ref{sec:ben-ent}.

\section{Quasiparticle picture for the charge-imbalance resolved fermionic negativity}
\label{sec:quasi-neg}

We now discuss the quasiparticle prediction for the charge-imbalance-resolved 
fermionic negativity. 
Here we focus on two intervals 
$A_1$ and $A_2$ embedded in an infinite chain. The two intervals 
are of the same length $\ell$, and are at distance $d$ (see Fig.~\ref{fig:cartoon}). 
We first determine the scaling behaviour of  
the charged moments of the partial time-reversed transpose $N_r(\alpha)$  
(cf.~\eqref{eq:nr-alpha}) after the quench from the N\'eel state. 
Here we only consider the case of the 
N\'eel quench because this allows us to exploit the results of Ref.~\cite{alba2022logarithmic}. 
To derive the quasiparticle prediction for $N_r(\alpha)$ and $N_1(\alpha)$  
we proceed as in section~\ref{sec:results}. To illustrate the idea, let us first consider 
the case of $\log(N_1(\alpha))$ (cf.~\eqref{eq:num}). The main simplification in the treatment of 
$N_1$ is that it depends only on the eigenvalues of the matrix $\Gamma_+$ (cf.~\eqref{eq:gammapm}). 
We can write $N_1(\alpha)$ as 
\begin{equation}
	\label{eq:toresum1}
	\log(N_1(\alpha))=-i\alpha+\sum_{j=0}^\infty c_{1,j}(\alpha)\mathrm{Tr}\,(\Gamma_+^j), 
\end{equation}
where the coefficients $c_{1,j}(\alpha)$ are obtained from 
\begin{equation}
	\label{eq:hra}
	h_{r,\alpha}(x)=\log \left[\left(\frac{1-x}{2}\right)^r+
		e^{i\alpha} \left(\frac{1+x}{2} \right)^r
	\right]=\sum_{j=0}^{\infty}c_{r,\alpha}(j)x^j, 
\end{equation}
with $r=1$. 
Here $\Gamma_\pm$ are the matrices defined in~\eqref{eq:gammapm}. 
Now, to determine the quasiparticle prediction for $\log(N_1(\alpha))$ one has 
to compute the hydrodynamic limit of $\mathrm{Tr}(\Gamma_+^j)$ for generic $j$. 
This has been computed in \cite{alba2022logarithmic}, and it reads

\begin{multline}
\label{eq:toresum}
\mathrm{Tr}(\Gamma_+^{2j})= 
2(1-a)^{2j}+[(b + 1 - a)^{2j} + (-b + 1 - a)^{2j}-2 (1 - a)^4] \Theta_1+\\
\frac{1}{2} \big[2 (b + 1- a) (-b + 1 - a))^j-2 (1 - a)^{2j}\big]  \Theta_2. 
\end{multline}
We also have 
\begin{equation}
	\label{eq:toresum2}
	\mathrm{Tr}(\Gamma_+^{2j+1})=0.
\end{equation}
In~\eqref{eq:toresum} and~\eqref{eq:toresum2}, $b=e^{-(\gamma^++\gamma^-)t}$ 
(cf.~\eqref{eq:C-map}) and we defined $a$ as 
\begin{equation}
	\label{eq:useful}
	a=2n_{\infty}(1-b)+b. 
\end{equation}
Here the functions $\Theta_1$ and $\Theta_2$ are defined as 
\begin{align}
	\label{eq:theta1}
\Theta_1=&\mathrm{max}(0,1-2|v(k)|t/\ell),\\
\label{eq:theta2}
\Theta_2=&\mathrm{max}(2|v(k)|t/\ell,2+d/\ell)+\mathrm{max}(2|v(k)|t/\ell,d/\ell)-2\mathrm{max}(2|v(k)|t/\ell, 1+d/\ell).
\end{align}
One has  $\Theta_1=1$ at $t=0$, and it decreases linearly up to $t=\ell/(2v(k))$. At longer  
times $\Theta_1$ is identically zero. On the other hand, $\Theta_2$ is zero for $t\le 
d/(2v(k))$. At larger times $\Theta_2$ grows linearly with time, up to $t=(d+\ell)/(2v(k))$,  
where it reaches a maximum. Then, $\Theta_2$ exhibits a linear decrease up to 
$t=(d+2\ell)/(2v(k))$. At longer times $\Theta_2$ is zero. The same function $\Theta_2$ 
describes the dynamics of the mutual information after quantum quenches without 
dissipation~\cite{coser2014entanglement,alba2019quantum}. 
By using~\eqref{eq:toresum},~\eqref{eq:toresum2} in~\eqref{eq:toresum1}, we obtain 
\begin{multline}
\label{eq:conj1}
\log (N_{1}(\alpha))=\ell\displaystyle 
\int_{-\pi}^{\pi}\frac{dk}{2\pi}
\mathrm{Re}\Big\{2h_{1,\alpha}+\Theta_1\big[h_{1,\alpha}(b+ 1 - a)+ 
	h_{1,\alpha}(-b + 1 - a) - 2h_{1,\alpha} (1 - a)\big]\\
+\frac{1}{2}\Theta_2\big[\widetilde{h}_{1,\alpha}((b + 1- a) (-b + 1 - a)) - 
2h_{1,\alpha} (1 - a)\big]\Big\}. 
\end{multline}
Here we introduced $\widetilde{h}_{r,\alpha}$ as 
\begin{equation}
\widetilde{h}_{r,\alpha}(x):=\log \left[e^{i\alpha}\frac{1+x+\cos(\alpha)(1-x)}{2} \right].
\end{equation}
%

%
\begin{figure}[t]
\centering
\includegraphics[width=0.58\textwidth]{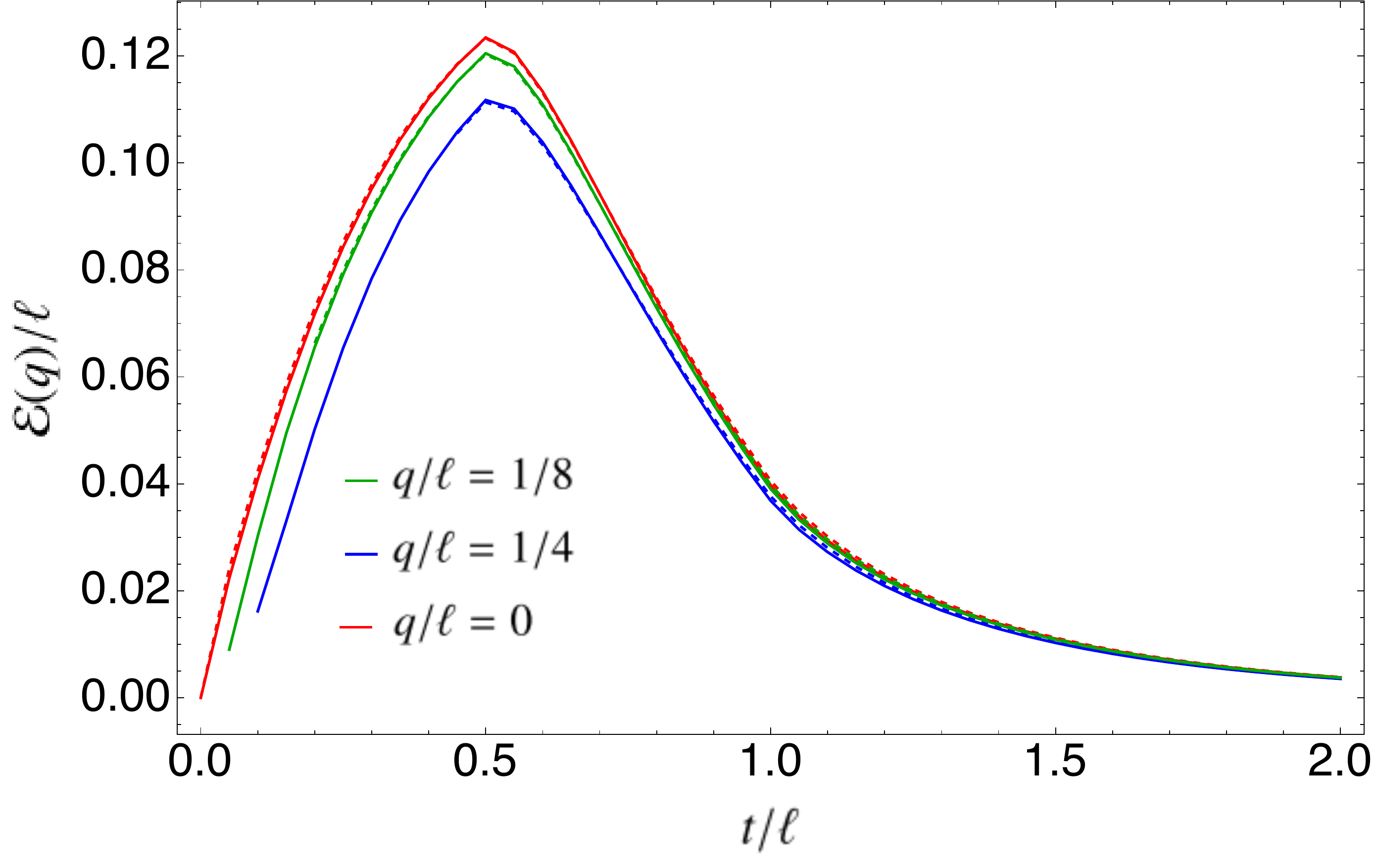}
\caption{ Effect of gain and loss dissipation in the charge-imbalance-resolved 
 negativity ${\cal E}(q)$: ${\cal E}(q)$ is plotted versus $t/\ell$.  
 Here we consider the negativity between two equal adjacent intervals of 
 size $\ell$. The data are for the quench from the N\'eel 
 state in the tight-binding chain. We consider several values of the 
 charge imbalance $q$. Our results are in the 
 weak-dissipative hydrodynamic limit, i.e., at fixed $t/\ell,q/\ell$ 
 and $\gamma^+=1/(2\ell)$ and $\gamma^-=1/(4\ell)$, with $\ell=80$ (continuous line) and $\ell=40$ (solid line). 
}
\label{fig:sr-neg-theo}
\end{figure}
%
To compute the charge-imbalance-resolved negativity, let us first 
consider the charged negativity ${\cal E}(\alpha)$ as 
\begin{equation}
	\label{eq:charged-neg}
	{\cal E}(\alpha)=\lim_{r\to 1}\log(N_{r}(\alpha)), \quad\mathrm{with}\,\,r\,\,
	\mathrm{even}, 
\end{equation}
where $N_r(\alpha)$ is defined in~\eqref{eq:nr-alpha}. 
The strategy, again, is to employ the results 
of~\cite{alba2022logarithmic}. 
Unlike the case in~\eqref{eq:conj1}, to evaluate $\log(N_r(\alpha))$ 
one has to obtain the behaviour of $\Gamma_\mathrm{x}^r$ (cf.~\eqref{eq:gammax}) 
for arbitrary $r$. These terms, however, can be obtained by using the results 
of Ref.~\cite{alba2022logarithmic}. 
The calculation is cumbersome and, since it is similar to Ref.~\cite{alba2022logarithmic}, 
we do not report it. For even $r$, the result reads as 
\begin{multline}
\label{eq:conj2}
\mathcal{E}_r(\alpha)=-i\alpha \ell+\frac{\ell}{2}\displaystyle 
\int_{-\pi}^{\pi}\frac{dk}{2\pi}\Big \{2h_{\frac{r}{2},\alpha}
\left(\frac{1}{2}
\pm \frac{1-a}{1+(1-a)^2} \right)+\\
\Theta_1\Big[h_{\frac{r}{2},\alpha}\left(\frac{1}{2}\pm 
	\frac{1-a-b}{1+(1-a-b)^2} \right)
	+
\big(b\to -b\big)
-2h_{\frac{r}{2},\alpha}\left(\frac{1}{2}\pm \frac{1-a}{1+(1-a)^2}  \right)\Big]+\\
\Theta_2\left[h_{\frac{r}{2},\alpha}\left(\frac{1}{2}\pm 
\frac{1-a}{[1+(1-a-b)^2]^{1/2}[1+(1-a+b)^2]^{1/2}}\right)
-h_{\frac{r}{2},\alpha}\left(\frac{1}{2}\pm \frac{1-a}{1+(1-a)^2} \right) \right]\Big\}\\
+r\ell \displaystyle \int_{-\pi}^{\pi}\frac{dk}
{2\pi}\Big\{ \left[2h_{2,0}\left(\frac{a}{2}\right)- 
h_{2,0}\left(\frac{a+b}{2} \right)- 
h_{2,0}\left(\frac{a-b}{2} \right)
\right]\Theta_3\\
+\left[	
h_{2,0}\left(\frac{a+b}{2} \right)+ h_{2,0}\left(\frac{a-b}{2} \right)\right] \Big\}, 
\end{multline}
where one has to sum over the $\pm$ signs. The functions $a$ and $b$ are the same as 
in~\eqref{eq:useful}. 
In~\eqref{eq:conj2} the function $h_{r,\alpha}$ is the same as in~\eqref{eq:hra}, whereas 
$\Theta_1$ and $\Theta_2$ are defined in~\eqref{eq:theta1} and~\eqref{eq:theta2}. 
In~\eqref{eq:conj2} we defined the new function $\Theta_3$ as 
\begin{multline}
\label{eq:theta3}
\Theta_3=2 \mathrm{min}(1/2,|v(k)|t/\ell)-2\mathrm{min}(0,1/2+d/(2\ell)-|v(k)|t/\ell) \\
+\mathrm{min}(0,1+d/(2\ell)-|v(k)|t/\ell)+
\mathrm{min}(0,d/(2\ell)-|v(k)|t/\ell). 
\end{multline}
Now, the first integral in~\eqref{eq:conj2} originates from the 
contribution of $\Gamma_\mathrm{x}$ (see the first term in~\eqref{eq:num1}), 
whereas the second 
one is due to the second term in~\eqref{eq:num1}. Notice that $\Theta_3$ is 
the same function describing the dynamics of the entropies of $A_1\cup A_2$ (see Fig.~\ref{fig:cartoon}). 
%
\begin{figure}[t]
\centering
\includegraphics[width=0.58\textwidth]{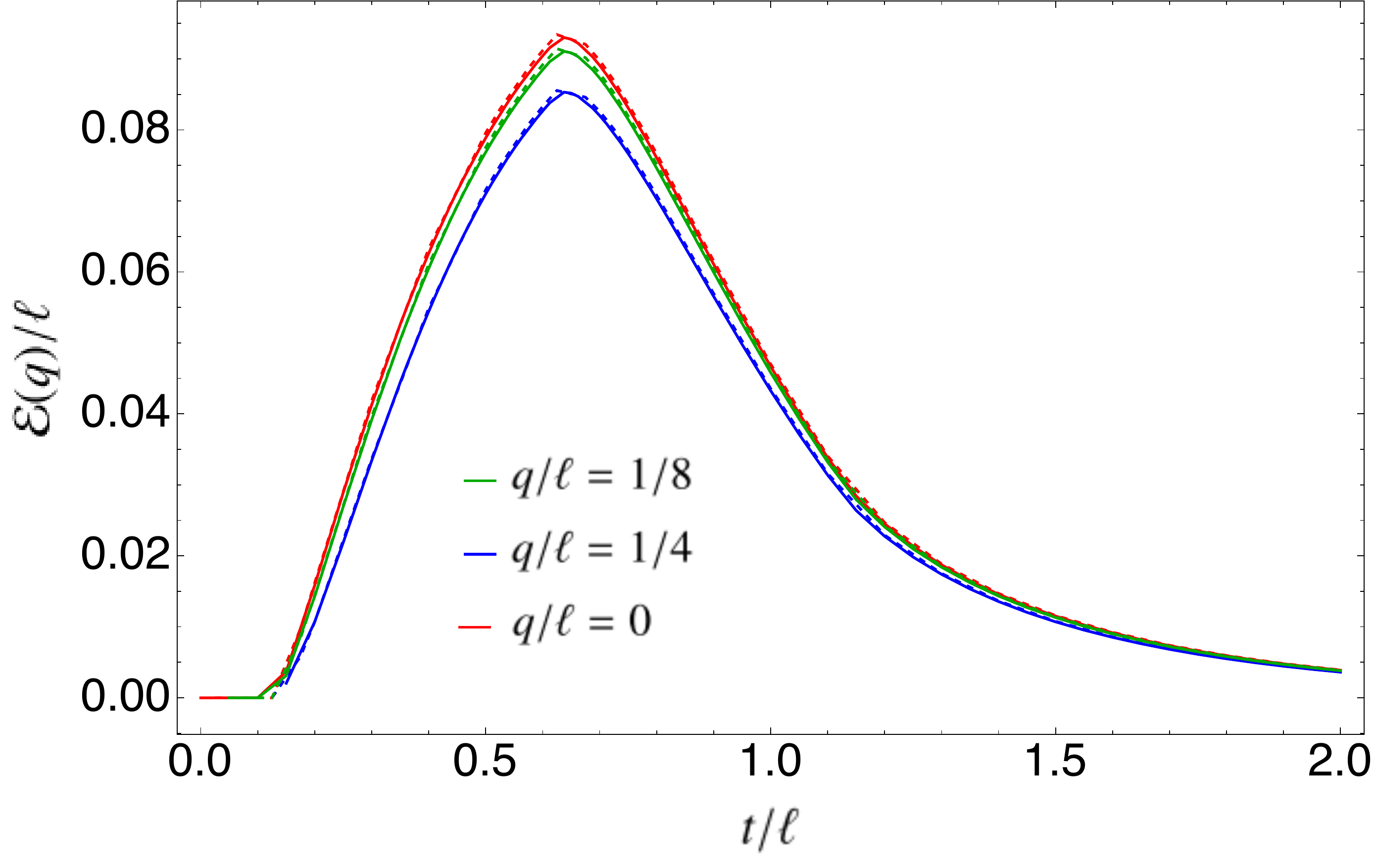}
\caption{ Same as in Fig.~\ref{fig:sr-neg-theo} for two disjoint 
 intervals at $d=\ell/4$. 
}
\label{fig:sr-neg-theo-a}
\end{figure}
%
The charged negativity is obtained by setting $r=1$ in~\eqref{eq:conj2}. 

From the expression for the charged negativity ${\cal E}_r(\alpha)$ (cf.~\eqref{eq:conj2}) 
we obtain the charge-imbalance-resolved negativity by first deriving 
the moments of the TR partial transpose ${\mathcal Z}_{R_1,r_e}(q)$ 
via the Fourier transform in~\eqref{eq:pqft}. 
This step has to be performed numerically. Finally, the charge-imbalance-resolved 
negativity ${\cal E}(q)$ is obtained from~\eqref{eq:rnq}. 

Our theoretical results for ${\cal E}(q)$ are shown in Fig.~\ref{fig:sr-neg-theo} and Fig.~\ref{fig:sr-neg-theo-a} for two equal-length adjacent and disjoint intervals, 
respectively. For two adjacent intervals we plot the density of negativity ${\cal E}(q)/\ell$ 
versus $t/\ell$. Our results hold in the weak-dissipative hydrodynamic limit. This corresponds 
to the usual hydrodynamic limit $t,\ell\to\infty$ with their ratio $t/\ell$ fixed. We also 
consider the limit $q\to\infty$ with fixed $q/\ell$. Finally, we consider the 
weak dissipation limit $\gamma^\pm\to0$ with fixed $\gamma^\pm \ell$. The charge-imbalance-resolved 
negativity exhibits a typical ``rise and fall'' dynamics. This behaviour is observed also 
in the absence of dissipation~\cite{alba2022logarithmic}. However, the short-time dynamics is not linear, in contrast with the non dissipative case. 
For each value of $q/\ell$ in Fig.~\ref{fig:sr-neg-theo} 
we show results for $\ell=40$ and $\ell=80$ (continuous and dashed line, respectively). The dependence  on $\ell$, which could arise because of the Fourier tranform, is quite weak. Finally, 
we observe that at very short times ${\cal E}(q)$ suffers from numerical instability 
in the evaluation of the Fourier transform for this geometry, 
therefore we do not show the plot for short times. While  it is likely that  
the presence of generic dissipation suppresses the time delay, as it happens for 
the entropies, clarifying this issue would require considerable numerical effort, 
and we leave it as an open problem. 
In Fig.~\ref{fig:sr-neg-theo-a} we consider two disjoint intervals at distance $d=\ell/4$. 
The behaviour of ${\cal E}(q)$ is similar to the case of adjacent intervals, the only 
difference being that now the negativity is zero up to $t\lesssim d/(2v_\mathrm{max})$, 
with $v_\mathrm{max}$ the maximum velocity in the system.

\section{Numerical benchmarks}
\label{sec:num}

In this section we provide numerical benchmarks for the results derived in 
sections~\ref{sec:results} and \ref{sec:quasi-neg}. Precisely, in section~\ref{sec:ben-mom} we discuss 
the charged moments, and in 
section~\ref{sec:ben-ent} and section~\ref{sec:qua} we focus on 
the symmetry-resolved entropy. In section~\ref{sec:log-corr} we 
investigate the weak dissipative limit and logarithmic scaling corrections. 
In section~\ref{sec:ben-neg} we 
consider the charge-imbalance-resolved negativity.

\subsection{Charged moments}
\label{sec:ben-mom}

%
\begin{figure}[t]
\centering
\subfigure
{\includegraphics[width=0.48\textwidth]{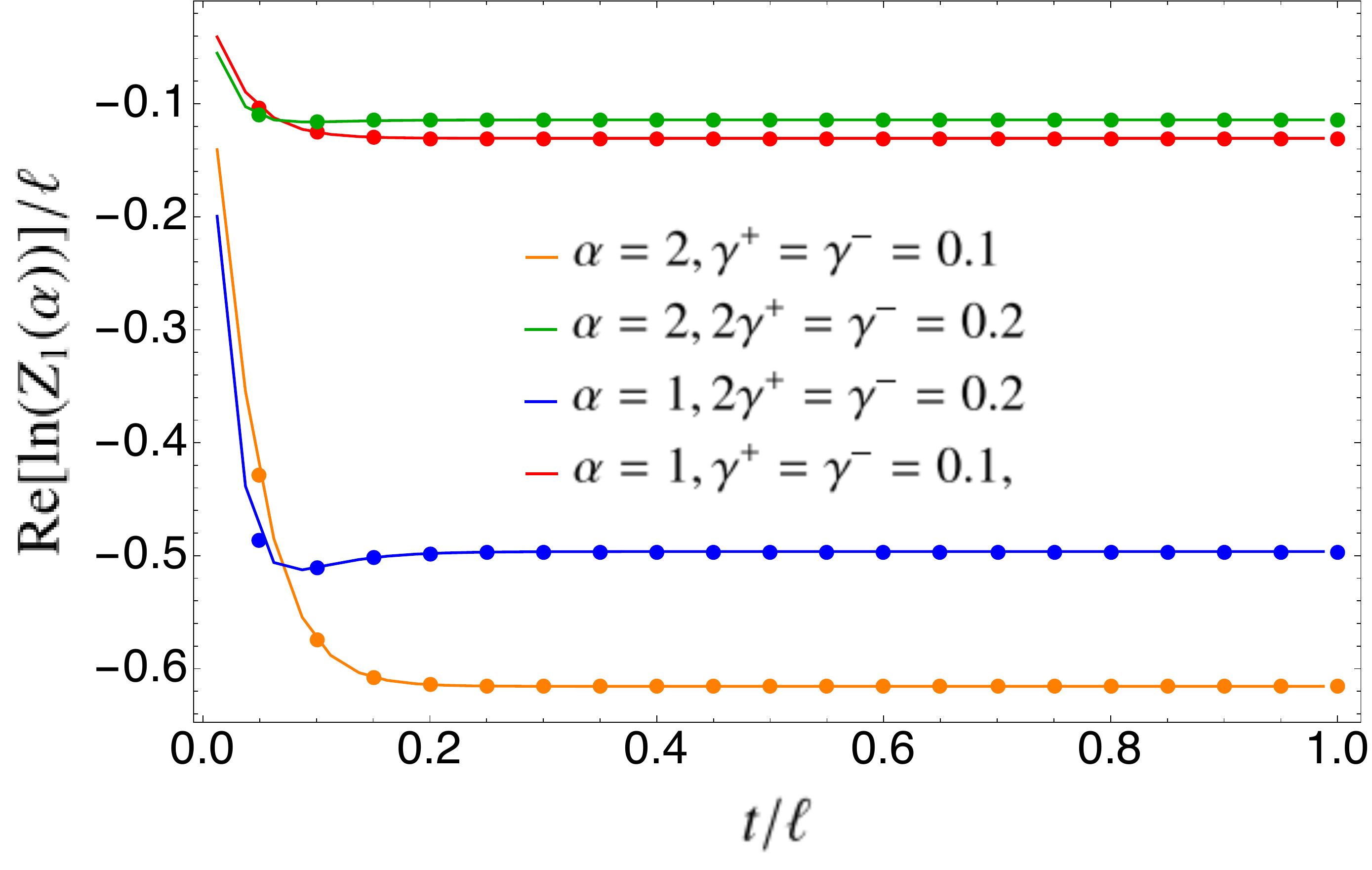}}
\subfigure
{\includegraphics[width=0.48\textwidth]{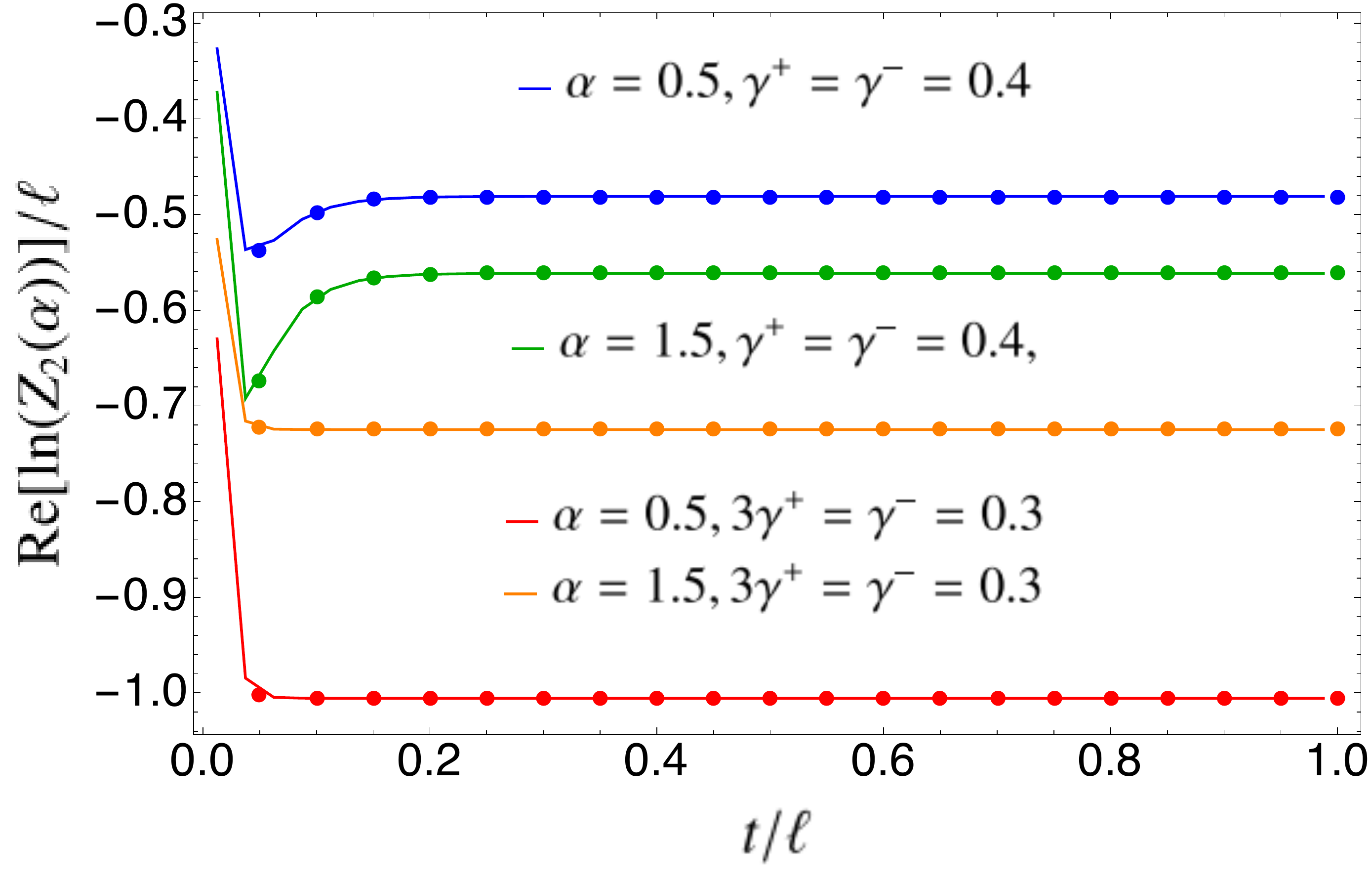}}
\caption{ Effect of gain/loss dissipation in the tight-binding chain after
 the quench from the fermionic N\'eel state: Real part of the logarithm of the
 charged moments $\mathrm{Re}[\log(Z_r(\alpha))]/\ell$ plotted versus $t/\ell$. 
 We show results for $r=1$ and $r=2$ in the left and right panel, respectively. 
 The symbols are exact lattice data for several values of $\alpha$ and gain/loss 
 rates $\gamma^\pm$. The data are for fixed $\ell=80$. 
 The continuous lines are the quasiparticle predictions (cf.~\eqref{eq:charged-quasi}).
 }
\label{fig:ZalphaN}
\end{figure}
%
%
\begin{figure}[t]
\centering
\subfigure
{\includegraphics[width=0.48\textwidth]{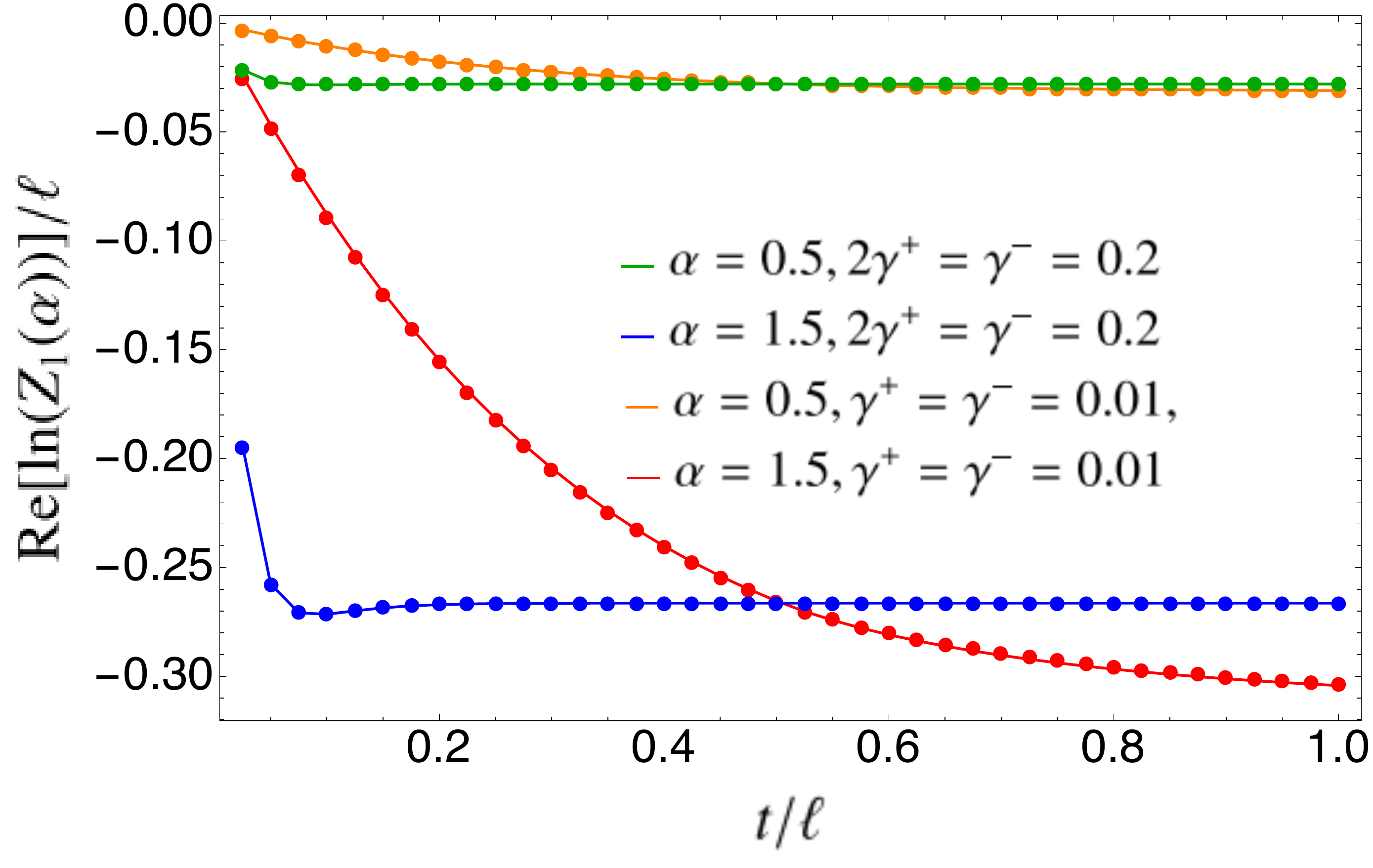}}
\subfigure
{\includegraphics[width=0.48\textwidth]{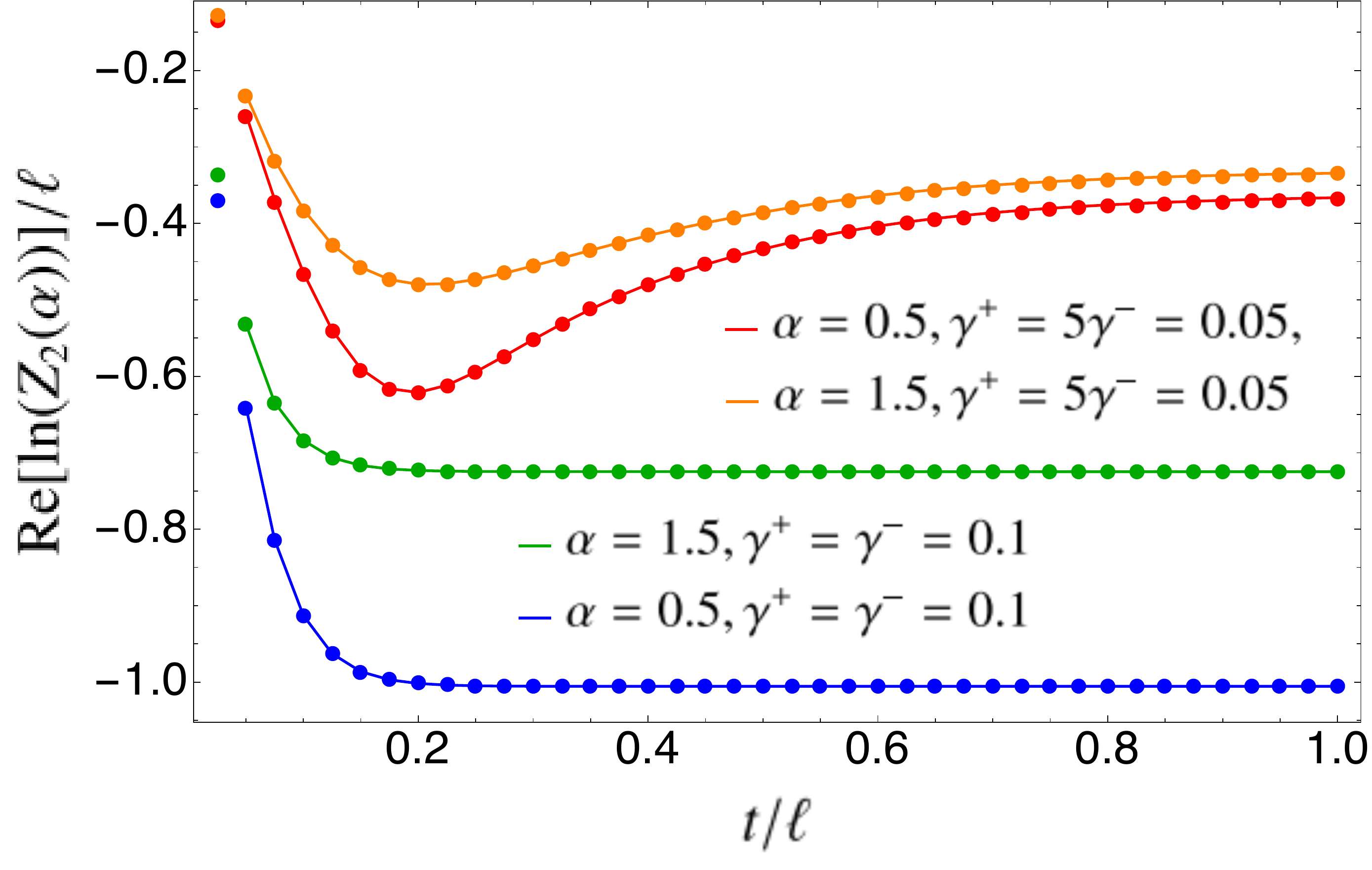}}
\caption{ Same as in Fig.~\ref{fig:ZalphaN} for the quench from the Majumdar-Ghosh 
 state. 
}
\label{fig:ZalphaD}
\end{figure}
%

Let us start discussing the hydrodynamic limit of the 
charged moments $Z_r(\alpha)$ (cf.~\eqref{eq:charged} for their 
definitions in the lattice model). We report numerical results for the 
charged moments in Fig.~\ref{fig:ZalphaN} and Fig.~\ref{fig:ZalphaD} for 
the quench from the N\'eel state and the Majumdar-Ghosh state, respectively. 
In the figures we show $\log(Z_r(\alpha))/\ell$ plotted versus $t/\ell$. 
The results are for a finite interval  of length $\ell$ embedded in the 
infinite chain. We show results for several values of the gain/loss rates 
$\gamma^\pm$ (symbols in the figure). 
Importantly, although we are interested in the weak-dissipative limit in which 
$\gamma^\pm\to0$ with $\gamma^\pm \ell$ fixed, here we show results for 
finite $\gamma^\pm$. The continuous lines in the figures is the theoretical 
prediction~\eqref{eq:charged-quasi}. Notice that at $t=0$ one 
has that 
\begin{equation}
	\label{eq:z-zero}
	\left.\log(Z_r(\alpha))\right|_{t=0}=\ell\int_{-\pi}^\pi\frac{dk}{2\pi}h^{\mathrm{mix}}_r
		(\alpha). 
\end{equation}
On the other hand, in the limit $t\to\infty$, one has that~\eqref{eq:charged-quasi} reduces to
\begin{equation}
	\label{eq:z-inf}
	\left.\log(Z_r(\alpha))\right|_{t\to\infty}=
		\ell\int_{-\pi}^\pi\frac{dk}{2\pi}h^{\mathrm{q}}_r(\alpha). 
\end{equation}
We should stress that since both $h^\mathrm{mix}_r(\alpha)$ and $h^\mathrm{q}_r(\alpha)$ 
depend on time, one has to take the limits $t=0$ and $t\to\infty$ also within the 
integrand in~\eqref{eq:z-zero} and~\eqref{eq:z-inf}. 
Now, as it is clear from Fig.~\ref{fig:ZalphaN} and Fig.~\ref{fig:ZalphaD}, 
the agreement between the numerical data (symbols in the figures) and the 
quasiparticle picture is perfect, even away from the weak dissipation limit. 

\subsection{Symmetry-resolved entropies}
\label{sec:ben-ent}

%
\begin{figure}[t]
\centering
\subfigure
{\includegraphics[width=0.48\textwidth]{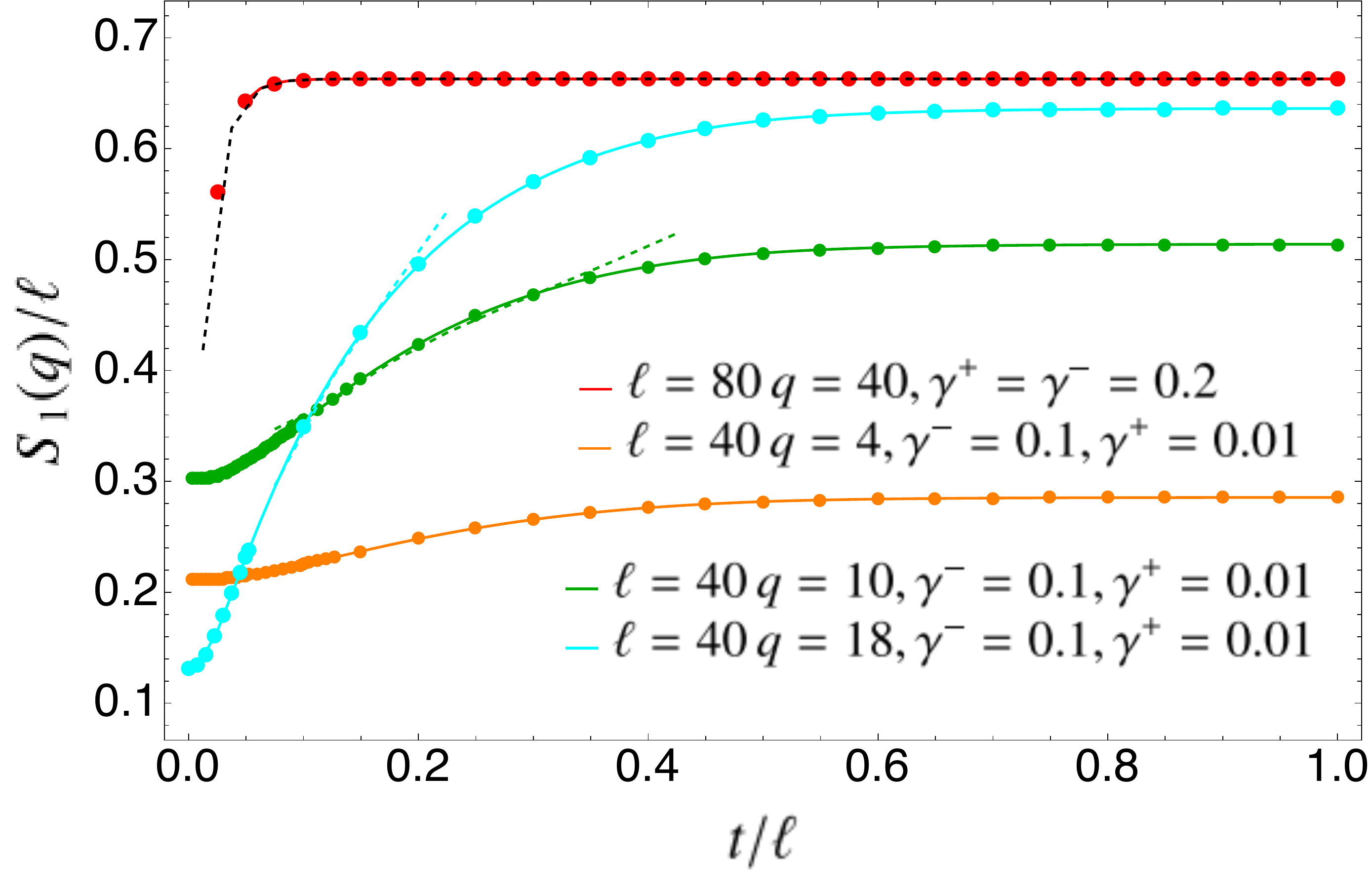}}
\subfigure
{\includegraphics[width=0.48\textwidth]{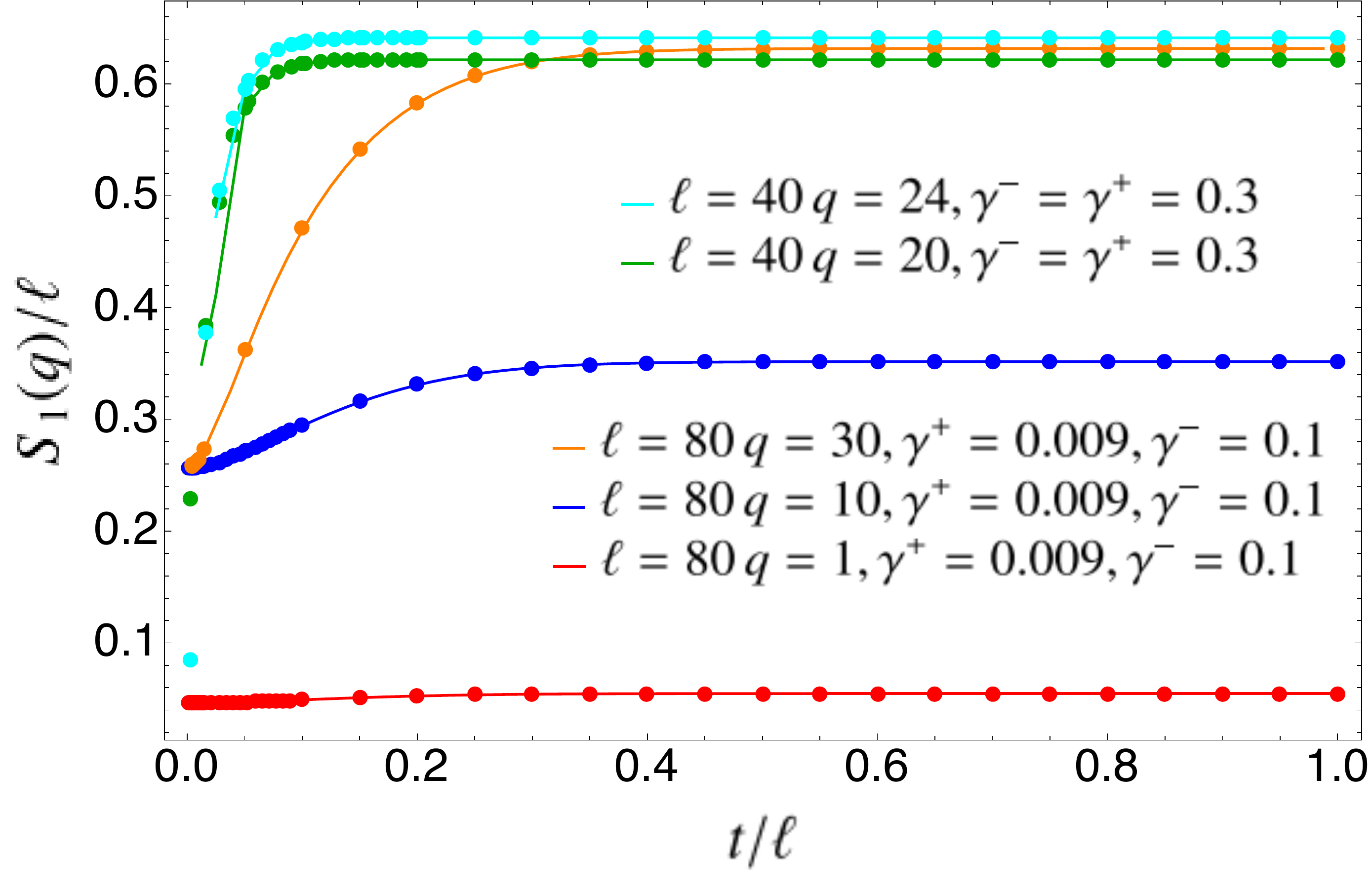}}
\caption{ Dynamics of the symmetry-resolved entanglement entropy $S_1(q)$ in the tight-binding 
 chain after the N\'eel (left) and dimer quench (right). 
 We show results for several values of $q$, interval size $\ell$, and dissipation 
 rates $\gamma^\pm$. The symbols are the lattice results. 
 The solid line is obtained by computing numerically Eq.~\eqref{eq:sqexact} as Fourier transform of the quasiparticle prediction \eqref{eq:charged-quasi}.
 The dashed line in the left panel is the quadratic approximation (cf.~\eqref{eq:sqapprox}). 
 }
\label{fig:sq}
\end{figure}
%

Having confirmed our theoretical results for the charged moments~\eqref{eq:charged-quasi}, 
we now consider the symmetry-resolved von Neumann entropy. As discussed in 
section~\ref{sec:quadratic}, the strategy is to plug the quasiparticle prediction~\eqref{eq:charged-quasi}
for the charged moments $Z_r(\alpha)$ in~\eqref{eq:FT}, then performing the Fourier transform 
numerically. In Fig.~\ref{fig:sq} we compare our analytic prediction against exact 
numerical data for the symmetry-resolved entropy. We show results for both the quench 
from the N\'eel state and the Majumdar-Ghosh state (left and right panel, respectively). We stress that the fact that the symmetry-resolved entropies are non-vanishing for $t>0$ for \textit{any} charge sector is due to the presence of dissipative terms: They suppress the delay and all the charge sectors are populated as soon as the state evolves, 
in contrast with what happens without dissipation~\cite{parez2021exact} or in the presence of pure loss/gain. 

%
\begin{figure}[t]
\centering
\subfigure
{\includegraphics[width=0.48\textwidth]{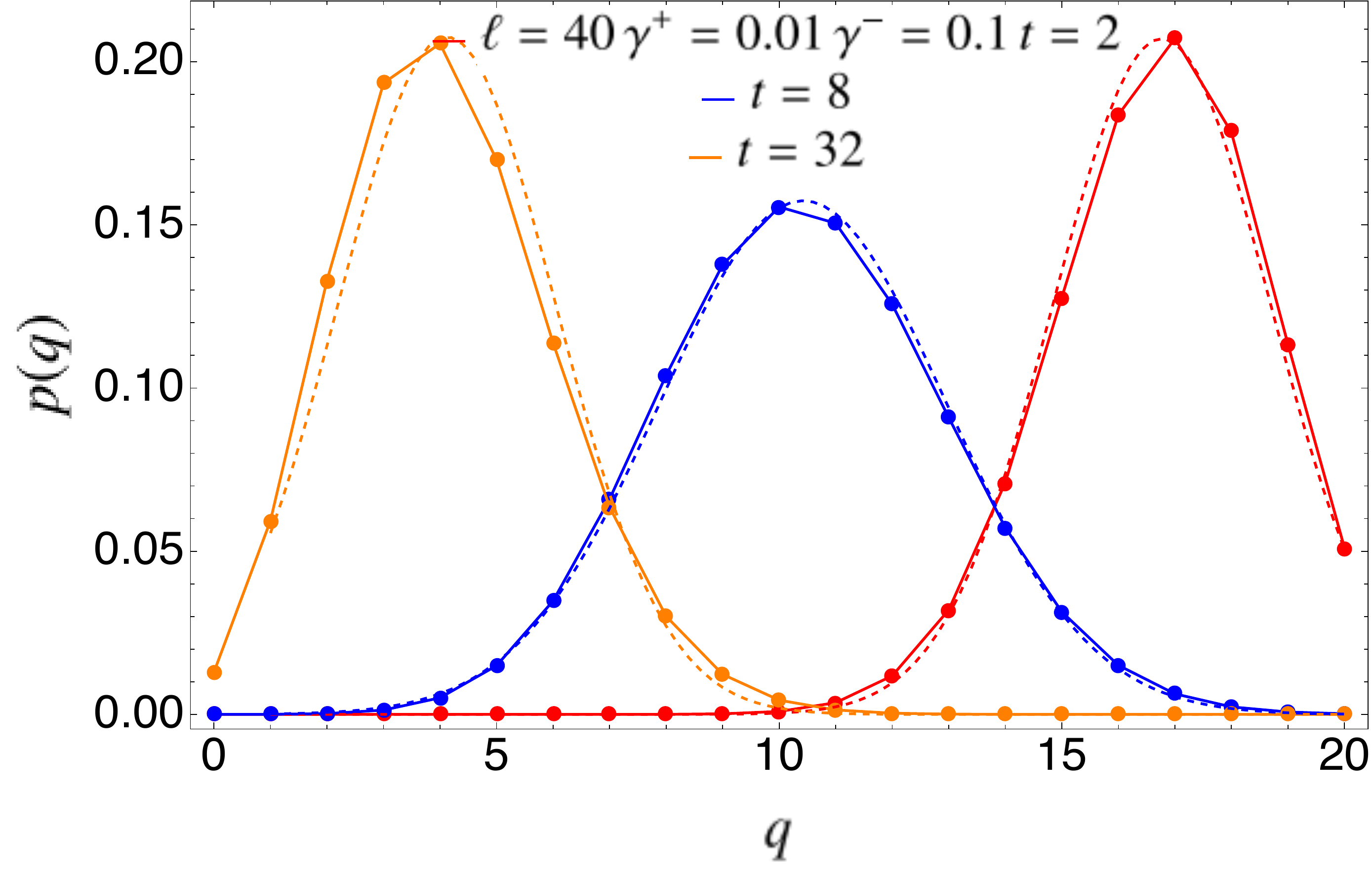}}
\subfigure
{\includegraphics[width=0.48\textwidth]{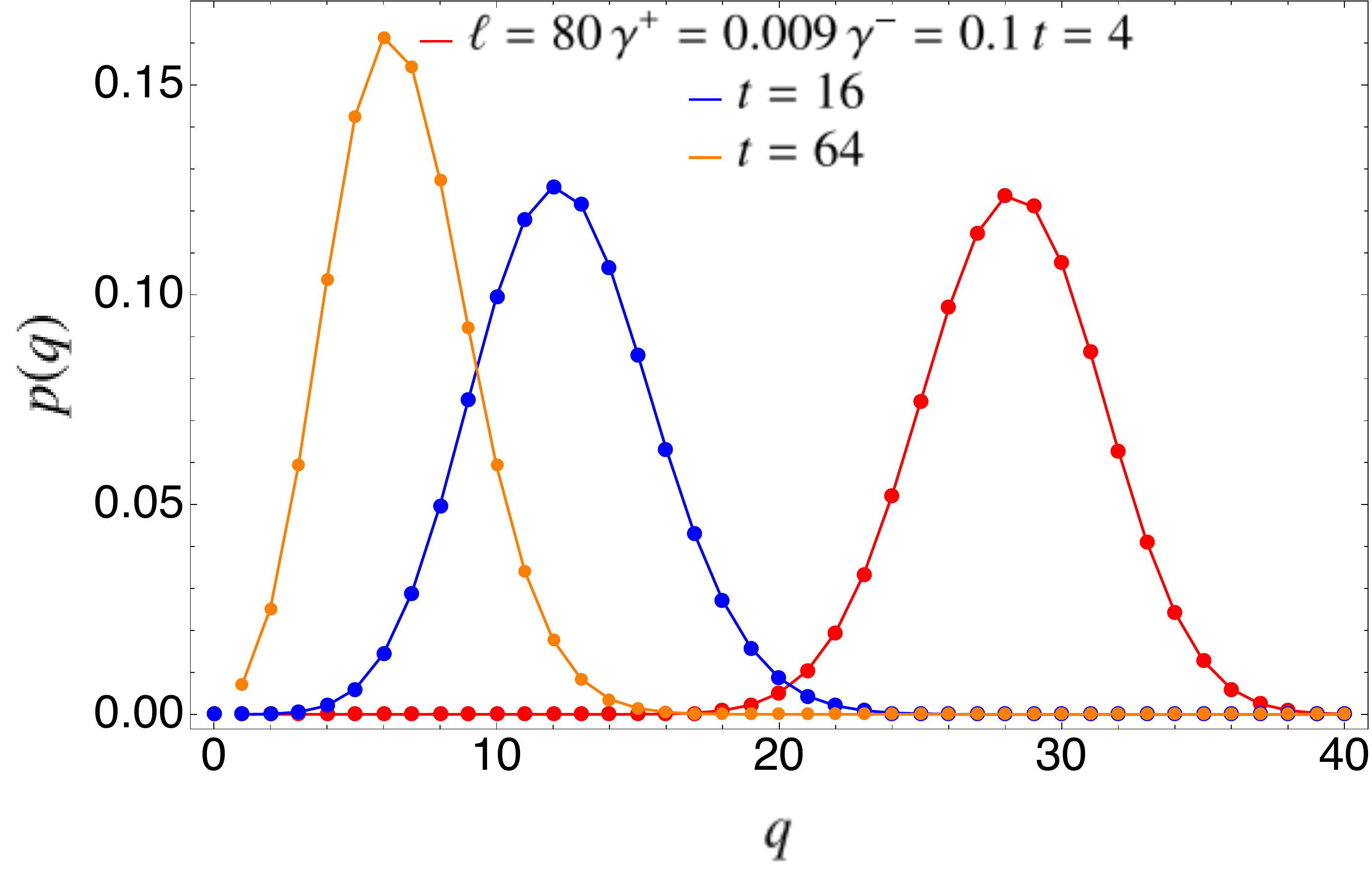}}
\subfigure
{\includegraphics[width=0.48\textwidth]{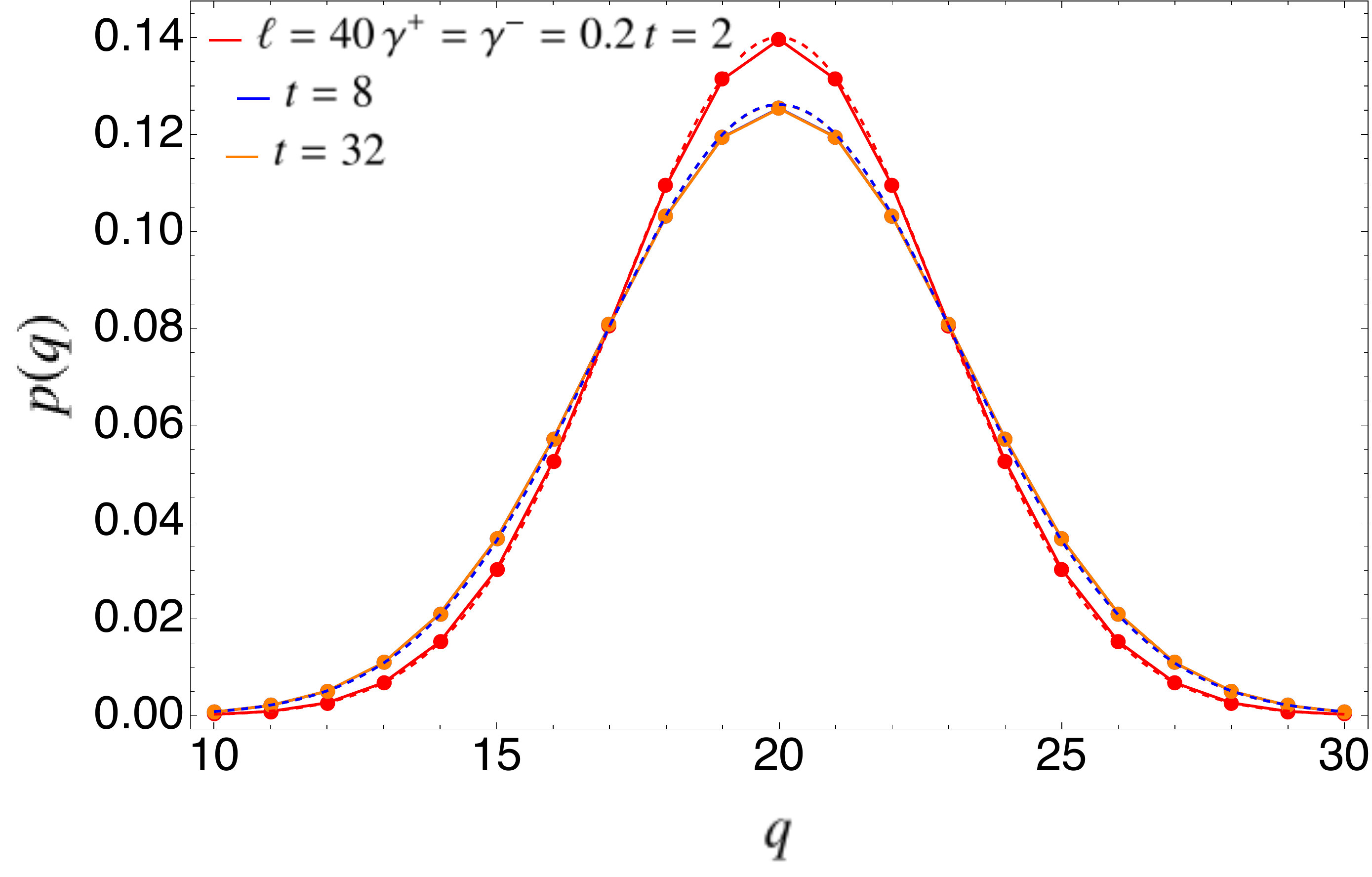}}
\subfigure
{\includegraphics[width=0.48\textwidth]{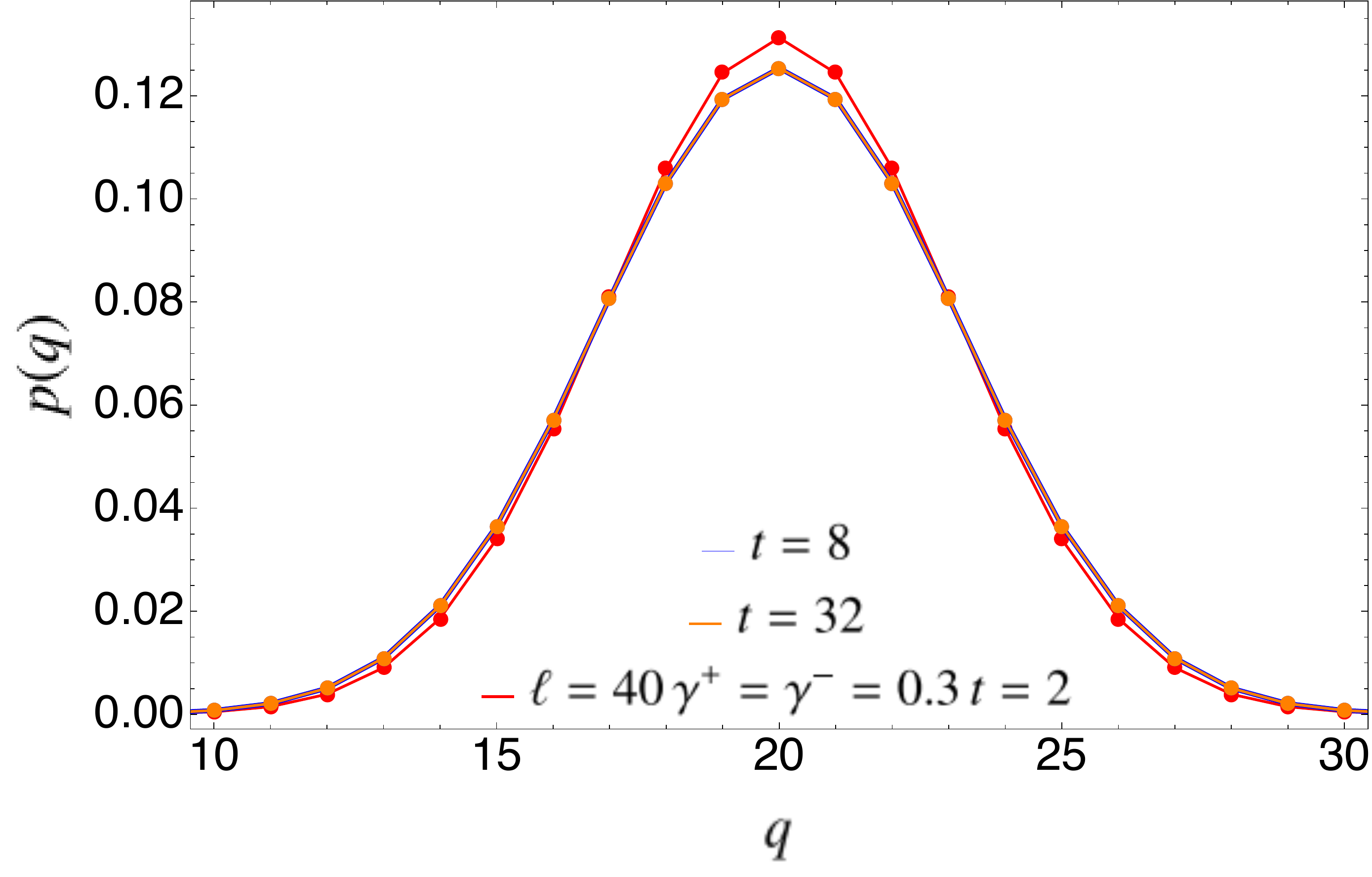}}
\caption{ Charge probability distribution of the tight-binding chain after
the N\'eel (left) or dimer quench (right) as a function of $q$ for fixed values 
of $\ell, t \gamma_{\pm}$. The solid line is obtained by performing numerically the integral in Eq. \eqref{eq:FT} as Fourier transform of the quasiparticle prediction \eqref{eq:charged-quasi}, while the dashed line for the N\'eel quench is 
obtained from the quadratic approximation (cf.~Eq.~\eqref{eq:pq}).
}
\label{fig:pq}
\end{figure}
%

\subsection{Quadratic approximation}
\label{sec:qua}

Let us now discuss the regime of validity of the quadratic approximation 
discussed in section~\ref{sec:quadratic}. This is obtained by expanding 
the charged moments $Z_r(\alpha)$ (cf.~\eqref{eq:charged}) around 
$\alpha=0$. Then, one can perform the Fourier transform in~\eqref{eq:FT} 
analytically. The result is reported in Fig.~\ref{fig:sq} (left) as dashed lines. 
Interestingly, the quadratic approximation works well for the case of 
balanced gain and loss dissipation, i.e., for $\gamma^+=\gamma^-$. 
However, away from the case of balanced dissipation the quadratic approximation 
fails to describe the dynamics of the entropy.

The reason of this failure is understood by monitoring the behaviour of the charge probability 
distribution $p(q)=\mathcal{Z}_1(q)$. We show results for $p(q)$ in 
Fig.~\ref{fig:pq}. The left and right panels in the figure are for 
the quench for the N\'eel and the Majumdar-Ghosh state, respectively, while top and bottom panels
show results for  $\gamma^+ \ll \gamma^-$ and $\gamma^-=\gamma^+ $. 
As we can notice from the figure,  the evolution of $p(q)$ is affected by the gain/loss terms. 
The main difference is that while for $\gamma^-=\gamma^+$ (i.e.  $n_{\infty} =1/2$), 
$p(q)$ remains peaked around $q=\ell/2$, for $\gamma^+\ll\gamma^-$,  
since $n_{\infty} \ll 1$ (cf.~\eqref{eq:C-map}), as $t$ increases, the peak of $p(q)$ moves  
toward smaller and smaller values of $q$.
Hence, since the symmetry resolved entropies in Fig. ~\ref{fig:sq} are for fixed value of $q$, 
for $\gamma^+\ll\gamma^-$ we have that $\Delta q=q-\bar q$ drifts as 
time passes to larger and larger values because of the drift of $\bar q$.
Now the quadratic approximation is valid only in the neighbourhood of $\Delta q=0$ (i.e. for $\Delta q  \ll \mathcal{B}^{(2)}_r$).  Working at fixed $q$, such condition breaks down unless $\gamma^-=\gamma^+$, for which 
$\bar q$ stays constant. 
This observation physically explains the failure of the quadratic approximation observed in Fig.~\ref{fig:sq}. 
Moreover, from Fig.~\ref{fig:sq} it is clear  
 that, for $\gamma_+=0.01, \gamma_-=0.1$, there is only a tiny time window (corresponding to $\Delta q\simeq 0$) where the quadratic approximation is valid. 

%
\begin{figure}[t]
\centering
{\includegraphics[width=0.58\textwidth]{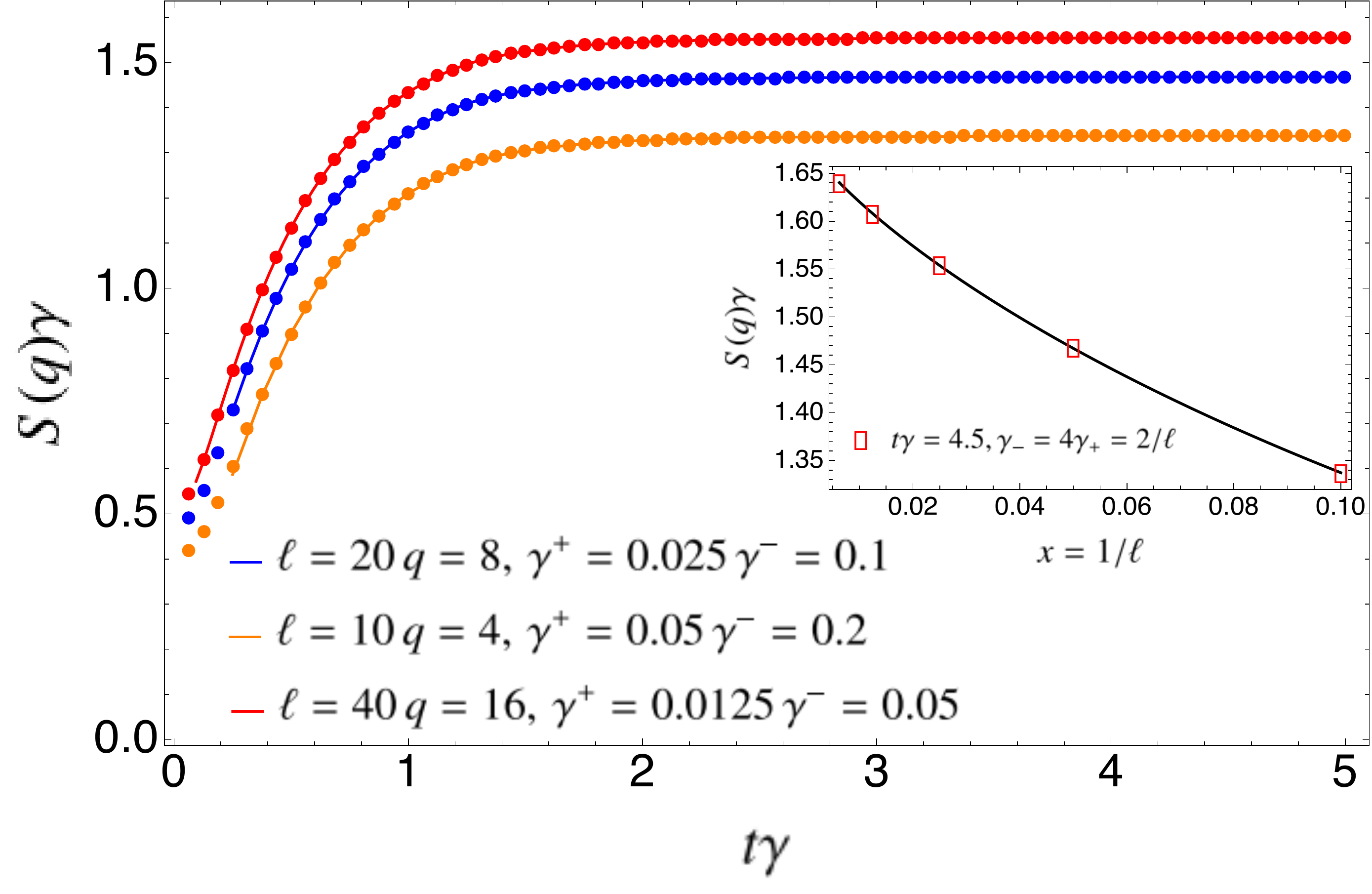}}
\caption{Symmetry-resolved entanglement entropy for the tight-binding 
 chain after the N\'eel quench in the  presence of gain and loss dissipation. 
 Weak-dissipative hydrodynamic scaling limit.
 We plot 
 $S(q)\gamma$ versus the rescaled time $t\gamma$, with $\gamma=\gamma^++\gamma^-$. 
 To reach the scaling limit we 
 rescale also the dissipation rates as $\gamma^+=1/(2\ell)$, $\gamma^-=2/\ell$. 
 Moreover, we fix $q/\ell=2/5$. Sizeable corrections 
 are visible and prevent from observing a data collapse. In the inset we 
 show data at fixed $t/\ell=1.8$ as a function of $1/\ell$. The continuous 
 line is a fit to $A-1/2\gamma \log(\ell)+B/\ell$, with $A,B$ fitting parameters. 
}
\label{fig:sq-collapse}
\end{figure}
%

\subsection{Logarithmic scaling corrections and the number entropy}
\label{sec:log-corr}

It is important to investigate explicitly the 
scaling behaviour in the weak-dissipative hydrodynamic limit. This is 
discussed in Fig.~\ref{fig:sq-collapse}. We plot exact numerical data for 
the symmetry-resolved von Neumann entropy for 
$\ell,t\to\infty$ at fixed $t/\ell$ and $\gamma^\pm\to0$ with 
fixed $\gamma^\pm\ell$. Moreover, we also fix $q/\ell$. Clearly, 
Fig.~\ref{fig:sq-collapse} shows that there are strong finite $\ell$ 
corrections. Indeed, in the weak-dissipative hydrodynamic limit all 
the data in Fig.~\ref{fig:sq-collapse} are expected to collapse on the 
same curve. Deviations from the expected behaviour 
can be attributed to the presence of logarithmic 
terms in $S(q)$. This is due to the fact that the result for 
the symmetry-resolved entropies should behave like (for $t \to \infty$)
\begin{equation}
	\label{eq:log}
	S_1(q)=a\ell-\frac{1}{2}\log \ell+O(1/\ell). 
\end{equation}
%
%
\begin{figure}[t]
\centering
{\includegraphics[width=0.58\textwidth]{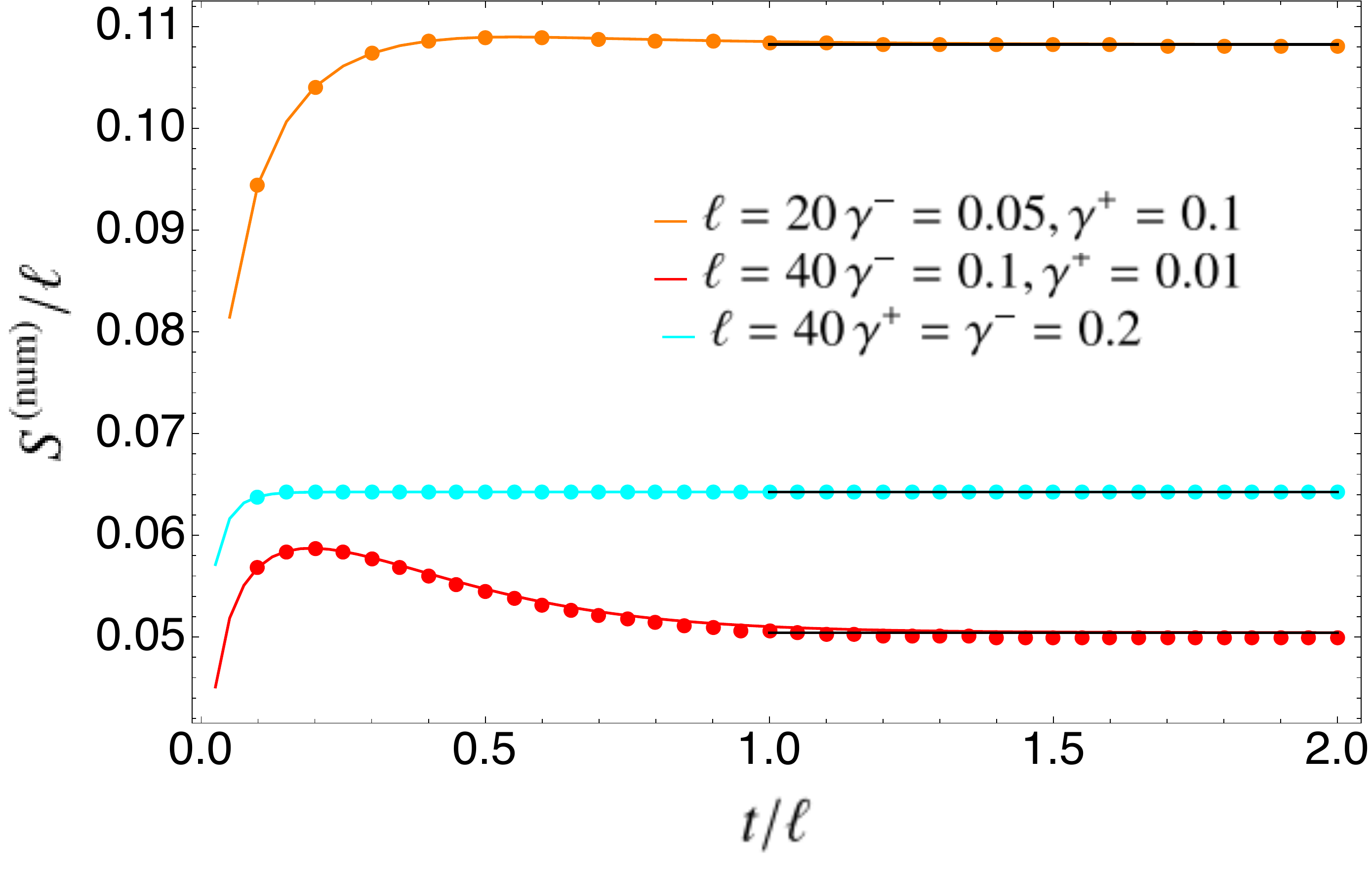}}
\caption{ Dynamics of the number entropy (cf.~\eqref{eq:decompositionSvN})
 $S^{(\mathrm{num})}$ after the quench from the N\'eel state in the tight-binding 
 chain with gain/loss dissipation. The symbols denote exact lattice 
 results, whereas the solid line is the prediction of the quasiparticle 
 picture (cf.~\eqref{eq:snum}). The horizontal line is the result in 
 the long time limit~\eqref{eq:larget}. 
 }
\label{fig:snum}
\end{figure}
%
The logarithmic correction in~\eqref{eq:log} is captured already by the 
quadratic approximation (cf.~\eqref{eq:S1-q}). 
In the sum rule~\eqref{eq:decompositionSvN}, this logarithmic term cancels with the number entropy and  so it can be better understood by studying the latter 
which is shown in Fig.~\ref{fig:snum}. 
The figure shows that at $t=0$ the number entropy is zero, reflecting that the initial state is  a product state. 
At later times the entropy grows, and it exhibits an exponential saturation at long times. 
The symbols in the figures are exact numerical data for $S^{(\mathrm{num})}$, 
whereas the continuous line is the analytic result obtained by using the 
quasiparticle picture. The result is obtained by using~\eqref{eq:charged-quasi-1} 
(see also~\eqref{eq:charged-quasi}), performing the Fourier transform~\eqref{eq:FT} 
numerically. The predictions of the quasiparticle picture are in perfect 
agreement with the lattice results, even away from the weak dissipative limit, i.e., 
for finite $\gamma^\pm$. In the figure we also show the long time limit~\eqref{eq:larget} 
(horizontal line), as obtained from the quadratic approximation, 
which describes quite well the lattice results.

\subsection{Charge-imbalance-resolved negativity}
\label{sec:ben-neg}

We now numerically investigate the validity of our results for the 
fermionic negativity (see section~\ref{sec:quasi-neg}). 
%
\begin{figure}[t]
\centering
\subfigure
{\includegraphics[width=0.48\textwidth]{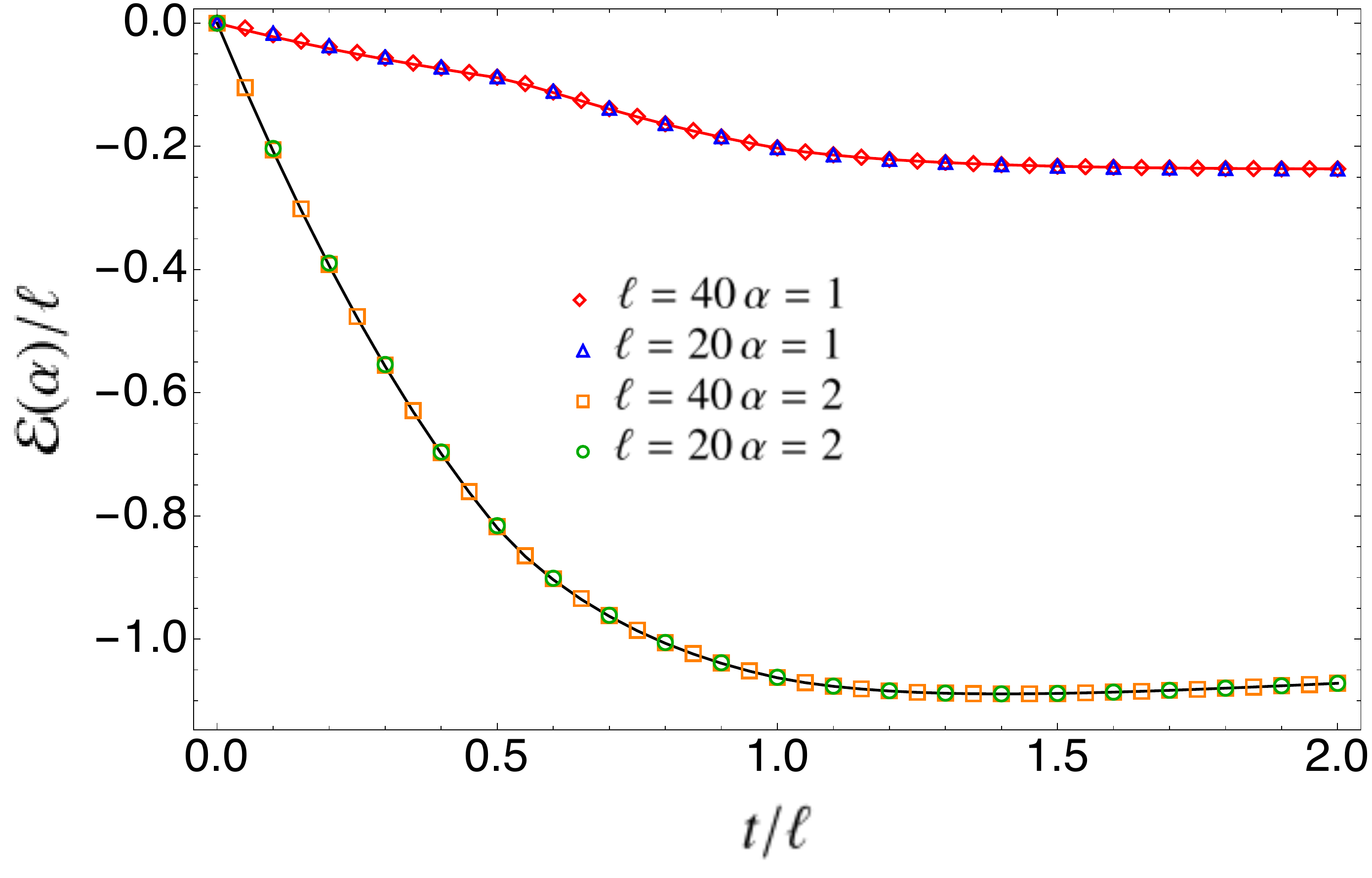}}
\subfigure
{\includegraphics[width=0.48\textwidth]{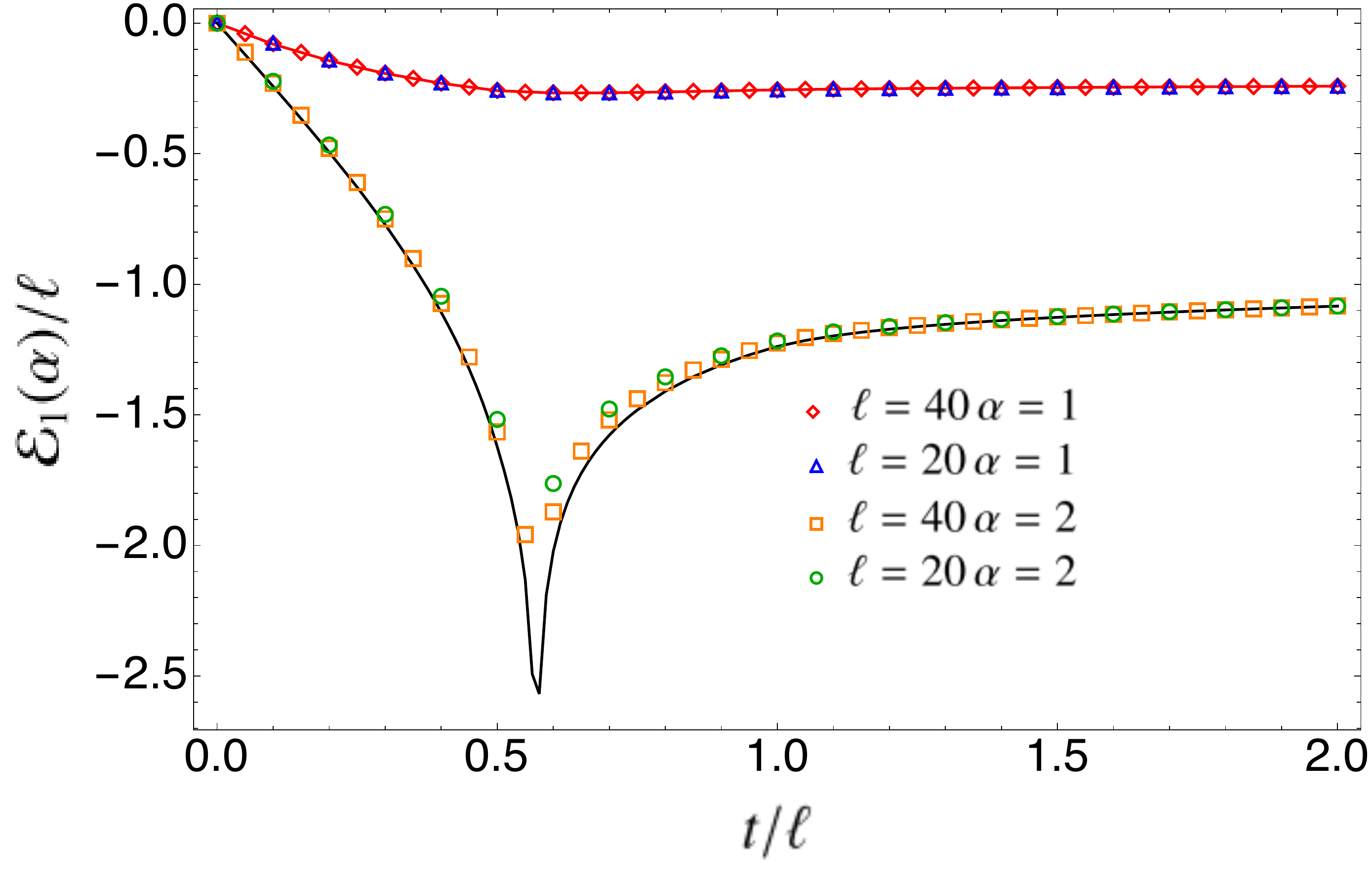}}
\subfigure
{\includegraphics[width=0.48\textwidth]{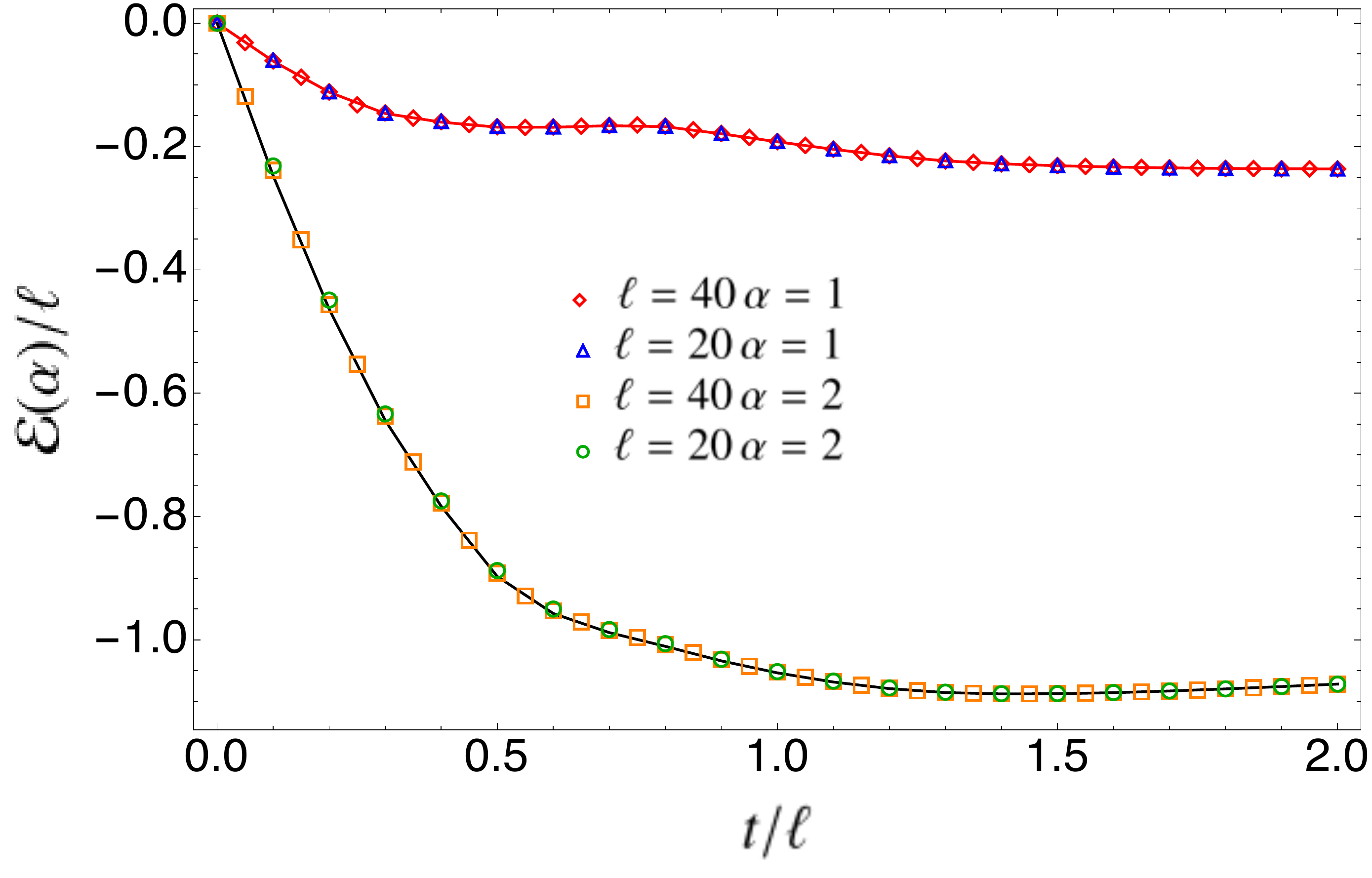}}
\subfigure
{\includegraphics[width=0.48\textwidth]{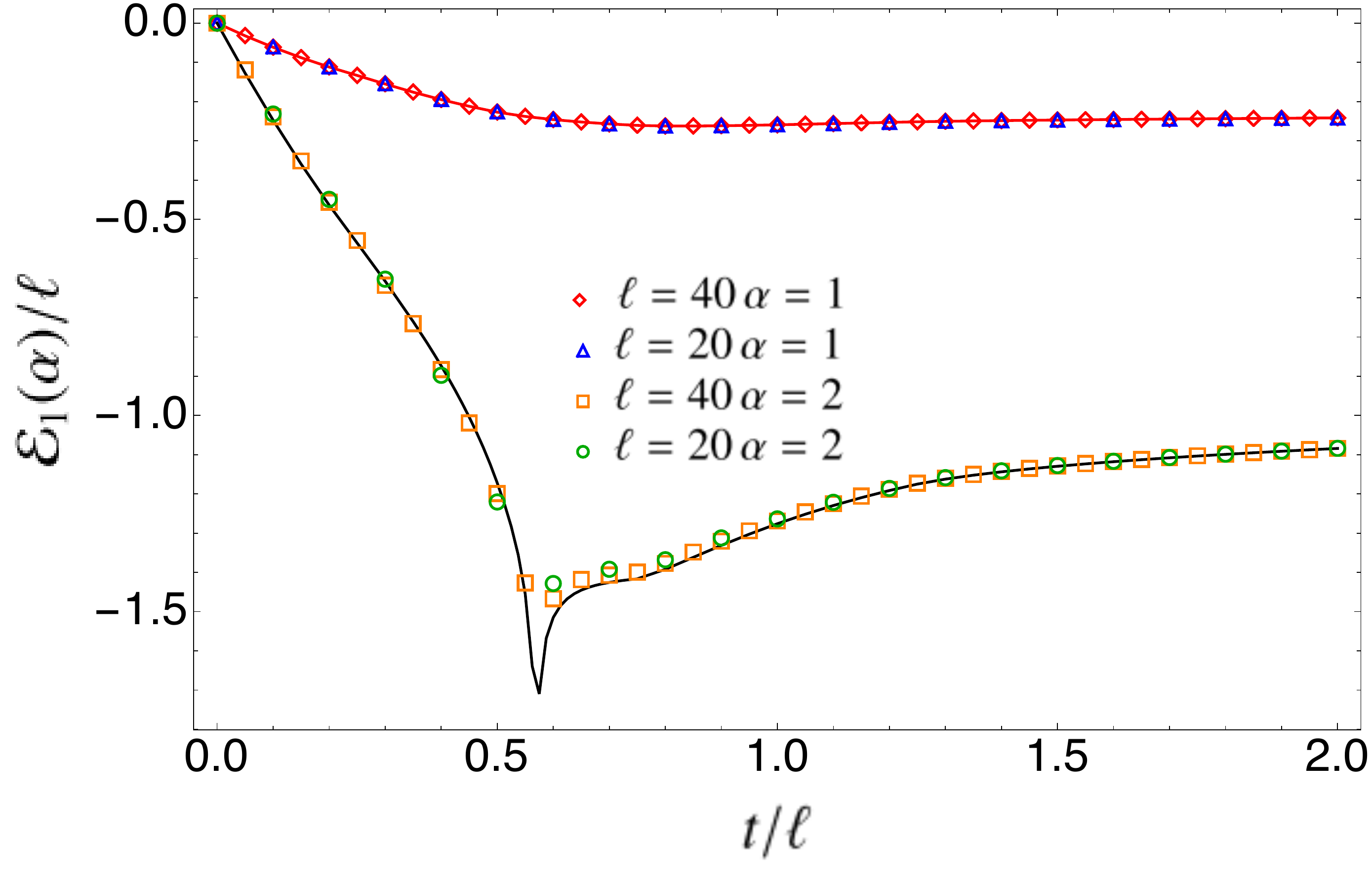}}
\caption{ Time evolution of $\mathcal{E}(\alpha)$ (left panels) and 
 $\mathcal{E}_1(\alpha)$ (right panels) after a quench from the N\'eel state
 in the tight-biding model. The top and bottom panels are for two adjacent 
 and two disjoint intervals at distance $d=\ell/2$, respectively 
 (see Fig.~\ref{fig:cartoon}). We plot the densities ${\cal E}(\alpha)/\ell$ 
 and ${\cal E}_1(\alpha)/\ell$ as a function of $t/\ell$. We fix 
 $\gamma_+=1/(2\ell)$, $\gamma_-=\gamma_+/2$. The data for different $\ell$ 
 at fixed $\alpha$ collapse on the same curve, which is in perfect 
 agreement with~\eqref{eq:conj2}  (left panels),  
 and~\eqref{eq:conj1}  (right panels).
}
\label{fig:neg2}
\end{figure}
%
%
\begin{figure}[t]
\centering
\subfigure
{\includegraphics[width=0.48\textwidth]{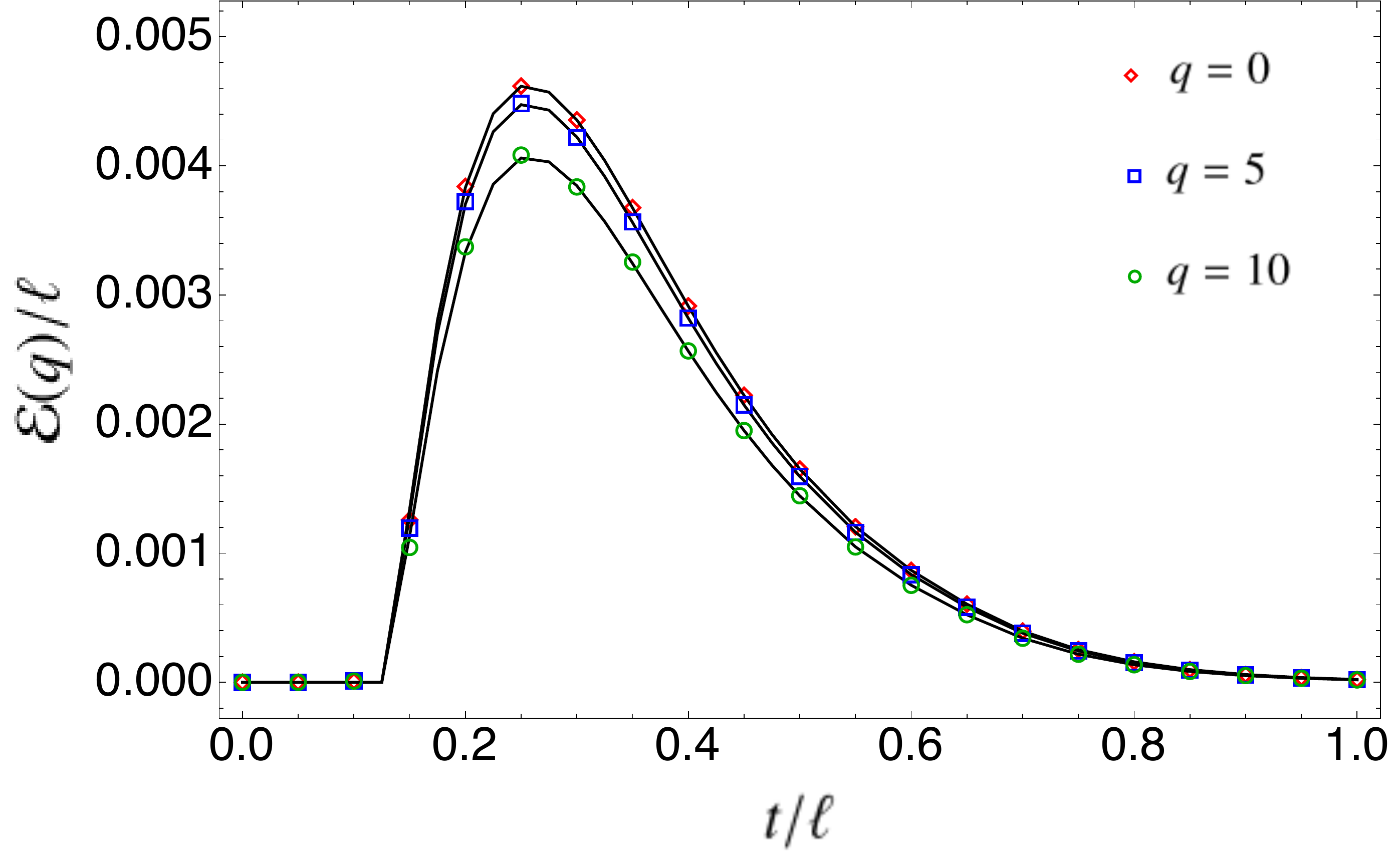}}
\subfigure
{\includegraphics[width=0.48\textwidth]{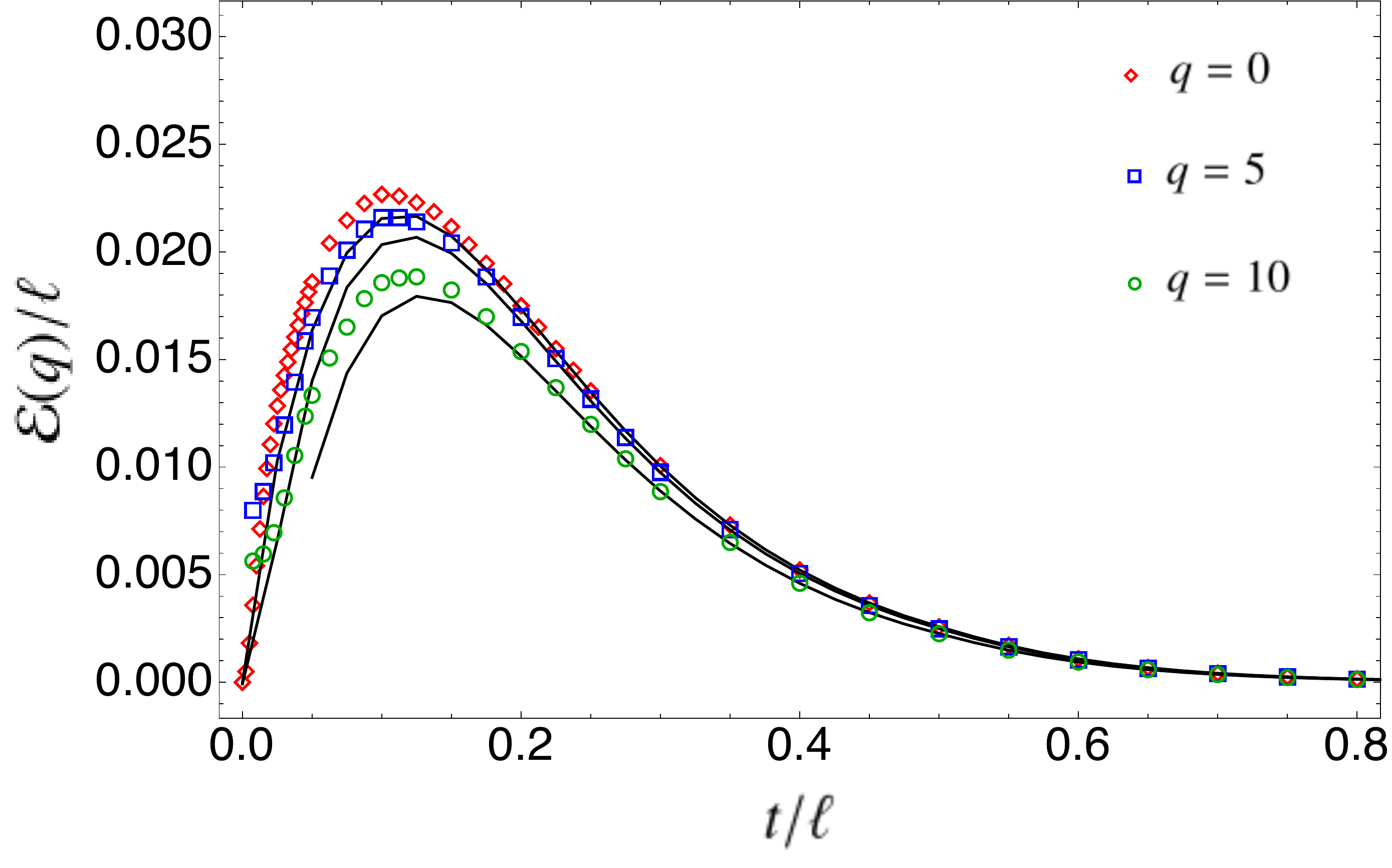}}
\subfigure
{\includegraphics[width=0.48\textwidth]{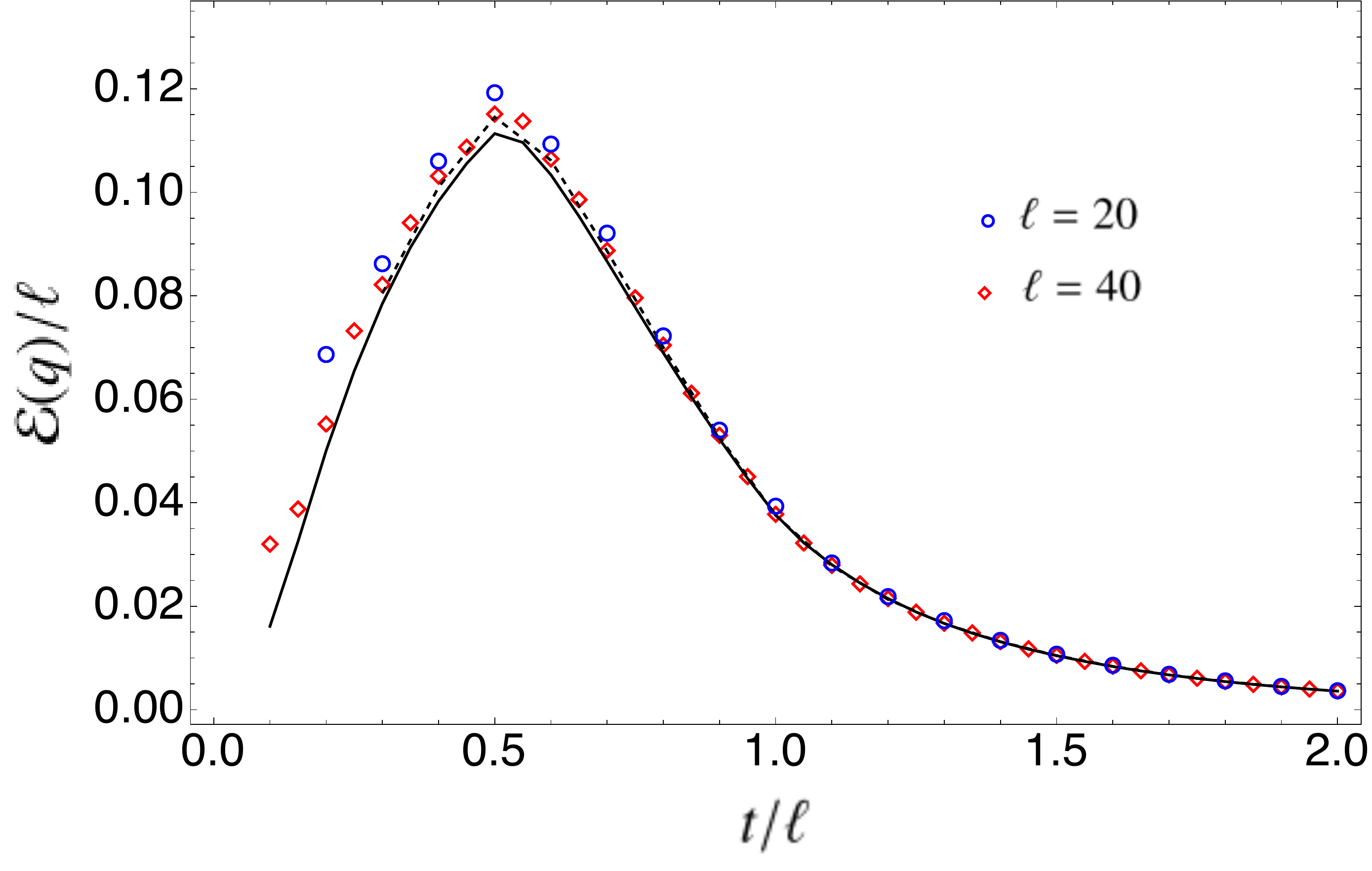}}
\caption{Top panel: 
 Time evolution of ${\mathcal{E}}(q)$ after a quench from the N\'eel state 
 in the tight-binding model  as a function of $t/\ell$ with $\ell=40$, 
 $d=\ell/4$ (left), $d=0$ (right) with fixed $\gamma_-=0.1$, $\gamma_+=0.01$. 
 Bottom panel: weak dissipative hydrodynamic limit.  
 ${\cal E}(q)$ for $\ell=20,40$, for $d=0$, 
 $\gamma_+=1/(2\ell)$, $\gamma_-=\gamma_+/2$, and $q/\ell=1/4$. 
}
\label{fig:imbalance}
\end{figure}
%
First, we consider the charged  
negativity ${\cal E}(\alpha)$ defined in~\eqref{eq:charged-neg}. 
In Fig.~\ref{fig:neg2} we show exact numerical data for ${\cal E}_1(\alpha)$ 
and ${\cal E}(\alpha)$ (left and right panels, respectively) 
in the tight-binding chain with gain and loss of fermions. We consider the quench from 
the fermionic N\'eel state. Since we are interested in the weak-dissipative hydrodynamic 
limit, we plot ${\cal E}(\alpha)/\ell$ versus $t/\ell$. To reach the weak-dissipation limit 
we rescale the dissipation rates $\gamma^\pm$ as $1/\ell$. Specifically, we choose 
$\gamma^+=1/(2\ell)$ and $\gamma^-=\gamma^+/2$. We show results for several values of $\alpha$. 
The top panels are for two adjacent intervals, whereas the bottom ones are for two disjoint ones 
at $d=\ell/2$. 
Our theoretical predictions  are~\eqref{eq:conj1} and~\eqref{eq:conj2}. 
Now, the symbols in Fig.~\ref{fig:neg2} are exact numerical data for 
${\cal E}(\alpha)$ and ${\cal E}_1(\alpha)$. Our theoretical 
predictions are shown as continuous lines in the figure, and are in perfect 
agreement with the numerical data. 

Finally, we discuss the charge-imbalance-resolved  negativity ${\mathcal{E}}(q)$. 
This is obtained by using~\eqref{eq:conj1} and~\eqref{eq:conj2} in~\eqref{eq:pqft}, 
performing the Fourier transform numerically, and using~\eqref{eq:rnq}. 
We show numerical results in Fig.~\ref{fig:imbalance}. The top panels in the figure 
are for fixed $\gamma^-=0.1$ and $\gamma^+=0.01$, $\ell=40$ and $q=0,5,10$. The 
top-right panel shows results for two adjacent intervals, i.e., for $d=0$, whereas 
the top-left one is for two disjoint intervals at $d=\ell/2$. The continuous 
lines in the figures are the theoretical results obtained from the 
quasiparticle picture. The agreement between lattice results and 
hydrodynamic limit is perfect for two disjoint intervals. Deviations from 
the scaling limit are visible for two adjacent intervals (top-right panel in 
Fig.~\ref{fig:imbalance}). 
These are due to the fact that our results are valid only in the 
weak-dissipative hydrodynamic limit. Similar corrections are observed 
in the absence of dissipation~\cite{alba2019quantum}. The right plot 
also clearly shows that there is no delay in the presence of dissipation, 
since $\mathcal{E}(q)$ is different from zero as soon as $t>0$, at least for $q=0,5$. 
The scaling behaviour for two adjacent intervals  
is better investigated in Fig.~\ref{fig:imbalance} in the bottom panel. We now 
show results in the limit $\ell,t,q\to\infty$ with $t/\ell,q/\ell$ fixed. 
Moreover, we consider $\gamma^\pm\to0$ at fixed $\gamma^\pm\ell$. 
Now, although for finite $\ell$ some deviations are 
present, upon increasing $\ell$ the numerical results 
approach the analytic predictions (lines in the figure).

\section{Conclusions}
\label{sec:concl}

We investigated the effects of gain and loss dissipation in the dynamics 
of the symmetry-resolved entropies and charge-imbalance-resolved 
negativity after a quantum quench in the tight-binding chain. This choice of dissipative terms preserves the block-diagonal structure of the reduced density matrix, and it is therefore suitable to study the symmetry resolution of entanglement in each charge sector.
We derived quasiparticle picture formulas for the dynamics of charged moments of the 
reduced density matrix, the symmetry-resolved von Neumann entropy, and the 
charge-imbalance-resolved negativity. Our results hold in the hydrodynamic limit of large 
intervals and long times, with their ratio fixed. Moreover, to ensure a nontrivial 
scaling behaviour we also considered weak dissipation. 
We showed that while the symmetry-resolved 
entropies are dominated by dissipative processes, the resolved negativity 
exhibits the typical rise and fall dynamics, reflecting that it is sensitive 
to entangled quasiparticles. Furthermore, we observed that the 
entropy does not exhibit a time delay (unless we consider a pure gain/loss evolution), contrarily to what happens in the non-dissipative case.

Let us now mention some possible directions for future research. First, 
it would be interesting to extend our results to other types of dissipation, 
for instance, for generic quadratic Lindblad master equations~\cite{carollo2022dissipative,alba2022hydrodynamics}. While this is 
feasible for the symmetry-resolved entropies, it is a challenging task for 
the negativity. Indeed,  even for the non-resolved negativity only the case of 
free fermions with gain and loss dissipation was investigated so far~\cite{alba2022logarithmic}.  
Clearly, it would be important to go beyond the case of quadratic Lindblad 
equations, considering more complicated dissipators, such as dephasing~\cite{wellnitz2022rise}.  
For instance, it would be interesting to employ recent results obtained for 
this dissipator in Ref.~\cite{coppola2023wigner}. Finally, one could study the effects of dissipative terms which do not preserve the block-diagonal structure of the reduced density matrix and quantify how much the symmetry is broken in this setup by using the entanglement asymmetry studied in \cite{amc-22-3,amvc-23}.
Similarly, one could consider initial states that explicitly breaks the internal symmetry and study the evolution of the asymmetry subject to gain and loss. 

\section*{Acknowledgments} 

We thank Gilles Parez and Vittorio Vitale for useful discussions. PC acknowledges support from ERC under Consolidator grant number 771536 (NEMO). 
SM thanks support from Caltech Institute for Quantum Information and Matter and the Walter Burke Institute for Theoretical Physics at Caltech.

\end{document}